\documentclass[prb,showpacs,preprintnumbers,amsfonts,amssymb,amsmath,floats,twocolumn,aps]{revtex4-1}

\usepackage{graphicx}%
\usepackage{bm}%
\usepackage{bbold}
\usepackage{color}%
\usepackage[caption=false]{subfig}
\renewcommand\thesubfigure{\alph{subfigure}}%

\usepackage[pdftex]{hyperref} %
\definecolor{darkblue}{rgb}{0,0,.5} %
\definecolor{black}{rgb}{0,0,0} %
\hypersetup{
  colorlinks=true, breaklinks=true, linkcolor=darkblue,
  menucolor=darkblue,
  urlcolor=darkblue, citecolor=darkblue }

\def \vs {\mathfrak} 
\def \op {\mathcal} 

\def \hop {J}%
\def \Hhop {\op H_\hop}%

\def \cre#1#2{\hat{#1}^\dagger_{#2}}%
\def \ann#1#2{\hat{#1}^{\phantom{\dagger}}_{#2}}%
\def \n#1#2{\hat n^{#1}_{#2}} \def \h#1#2{\hat h^{#1}_{#2}}%

\def \cc#1{\cre{c}{#1}}%
\def \ca#1{\ann{c}{#1}}%
\def \nc#1{\n{c}{#1}}%
\def \hc#1{\h{c}{#1}}%

\def \green#1#2{\langle\negthinspace\langle#1;#2\rangle\negthinspace\rangle}%

\def \ua{\uparrow}%
\def \da{\downarrow}%
\def \ra{\rightarrow}%
\def \com[#1,#2]{\left[#1,#2\right]}

\def \ket|#1>{\left|#1\right>}

\def\ff#1{\boldsymbol #1}

\begin{document}
  
\title{Doublon dynamics in the extended Fermi Hubbard model}

\author{Felix Hofmann}
\email{fhofmann@physik.uni-hamburg.de}
\affiliation{I. Institut f\"ur Theoretische Physik, Universit\"at
  Hamburg, Jungiusstra\ss{}e 9, 20355 Hamburg, Germany}
\author{Michael Potthoff}
\affiliation{I. Institut f\"ur Theoretische Physik, Universit\"at
  Hamburg, Jungiusstra\ss{}e 9, 20355 Hamburg, Germany}

\begin{abstract}
  Two fermions occupying the same site of a lattice model with
  strongly repulsive Hubbard-type interaction $U$ form a doublon, a
  long-living excitation the decay of which is suppressed because of
  energy conservation.  By means of an exact-diagonalization approach
  based on the Krylov-space technique, we study the dynamics of a
  single doublon, of two doublons, and of a doublon in the presence of
  two additional fermions prepared locally in the initial state of the
  extended Hubbard model.  The time dependence of the expectation
  value of the double occupancy at the different sites of a large
  one-dimensional lattice is analyzed by perturbative arguments.  
  In this way the spatiotemporal evolution of the doublon can be understood.
  The initial decay takes place on a short time scale $1/U$, and the long-time 
  average of the decayed fraction of the total double occupancy scales as $1/U^2$.
  We demonstrate how the dynamics of a
  doublon in the initial state is related to the spectrum of
  two-fermion excitations obtained from linear-response theory, we
  work out the difference between doublons composed of fermions vs
  doublons composed of bosons, and we show that despite the increase of
  phase space for inelastic decay processes, the stability of a doublon
  is enhanced by the presence of additional fermions on an
  intermediate time scale.     
\end{abstract}
 
\pacs{71.10.Fd, 67.85.-d, 78.47.D-}

\maketitle

\section{Introduction}
\label{sec:introduction}

Since the seminal work of Jaksch \emph{et al.},\cite{jaksch1998cold}
ultracold atomic gases in optical lattices have served as a valuable
testing ground for the rich phenomenology of many-body models which
originally were introduced in the context of condensed-matter
physics.\cite{bloch2008manybodyultracoldgases,trefzger2011ultracolddipgasesoptlatt,lewenstein2007ultracoldatomicgases}
A nice example is the concept of repulsively bound pairs of fermions
which can be studied in the strong-coupling regime of the Hubbard
model or, as shown
recently,\cite{wall2011quantuminterferencechargeexpaths} in an organic
salt at room temperature by means of ultrafast optical spectroscopy.
Repulsively bound pairs, named doublons, are already known since the
early work of Hubbard\cite{hubbard1963electron} and were lately
addressed in both theoretical and experimental work in
bosonic\cite{winkler2006repulsivelybound,petrosyan2007quantumliquid,petrosyan2008twopartstateshubbardmodel,valiente2008quantumdynonetwobosonic,valiente2009twopartboundstateexthubbard,javanainen2010dimertwobosons1doptlatt}
as well as
fermionic\cite{jochim2003becmolecules,greiner2003emergencemolecularbecfermigas,rosch2008metastable,hassanieh2008excitonsin1dhubbard,heidrichmeisner2009quantdistillation,strohmaier2010observationdoublondecay,hansen2011splithubbardbands,enss2011lightconerenorm,kajala2011xepdyn1dfermihubbardmodel}
Hubbard-type models.  The fermionic case directly refers to
condensed-matter systems, such as strongly correlated electrons in a
valence band of transition-metals and their compounds, and
two-particle electron spectroscopy.

A doublon is a pair of two fermions tightly bound to each other.  The
pair is itinerant, it propagates through the lattice and thereby
acquires a certain energy dispersion.  The pair may decay into its
constituents.  However, for strongly repulsive interaction $U>0$, this
decay is suppressed very efficiently.  The stability of the doublon
appears as counterintuitive since an energy of the order of $U>0$
would be gained if the two fermions were propagating through the
lattice independently.  There is, however, a ``repulsive binding''
originating from energy conservation: For $U$ much larger than the
nearest-neighbor hopping $J$, the excess energy $U$ cannot be
accommodated in the kinetic energy of the two independent fermions
which at most amounts to twice the bare bandwidth $W \propto J$.

In the strong-coupling limit, doubly occupied sites are created in
different types of electron spectroscopies:\cite{potth2001theor} The
spectral function $A_{ij}(\omega)$ at positive frequencies $\omega>0$,
obtained from the imaginary part of the one-particle Green's function
$\green{\ca{i\sigma}}{\cc{j\sigma}}$, is related to inverse
photoemission, and the upper Hubbard band in the spectral function
describes a final state with doubly occupied sites.  The lower Hubbard
band represents the analog of the upper Hubbard band in case of
photoemission.  For the Hubbard model on a bipartite lattice at half
filling, it is obtained from the upper one by a particle-hole
transformation and thus describes repulsively bound holes. Doublons
can also be created in an otherwise empty valence band in a
two-particle process, such as appearance-potential spectroscopy (APS),
i.e., the ``time inverse'' of Auger-electron spectroscopy (AES).
Here, two additional electrons (holes) are created, preferably at the
same site, in the final state of APS (AES).  Doublon bound states in
APS/AES show up in the local two-particle Green's function
$\green{\ca{i\sigma}\ca{i\bar\sigma}}{\cc{i\sigma}\cc{i\bar\sigma}}$
as is well known from Cini-Sawatzky
theory.\cite{cini1977two,sawatzky1977quasiatomic,nolting1990influence,potth2001theor}
Furthermore, doublons appear in particle-hole excitations associated
with Green's functions of the type
$\green{\cc{i\sigma}\ca{j\sigma}}{\cc{k\sigma}\ca{l\sigma}}$.  In all
mentioned cases, a doublon would be identified with a long-lived
excitation at energies of the order of $U$.

Since a pair of fermions has bosonic character, the exciting question
arises whether a macroscopically large number of doublons could Bose
condensate at sufficiently low temperatures and high densities.  This
has been studied theoretically for doublons of bosonic
\cite{petrosyan2007quantumliquid} and of fermionic
constituents.\cite{rosch2008metastable} The two main questions in this
context concern the doublon stability and the effective interaction
between doublons: First, a sufficiently long lifetime of doublons is
required for a possible Bose condensation taking place in a metastable
state.  Recent experiments with fermionic atoms in optical lattices
\cite{strohmaier2010observationdoublondecay} in fact give a lifetime
which increases exponentially with $U$.  Second, the physical
interactions between the constituents give rise to an effective
doublon-doublon interaction in an effective low-energy theory.  A
strong repulsive $U$, for example, leads to an effectively attractive
interaction between doublons formed by two bosons.  It has been shown
that this inhibits condensation but rather favors phase
separation.\cite{petrosyan2007quantumliquid}

The real-time dynamics of a spatially extended system of strongly
correlated fermions poses a notoriously complex many-body problem
which is hardly accessible to exact analytical or numerical methods.
Either one has to tolerate mean-field-type approximations like in the
nonequilibrium dynamical mean-field approach
\cite{schmidt2002noneqDMFTstrongcorrsys,freericks2006noneqDMFT,eckstein2009thermalization}
or has to restrict oneself to one-dimensional or impurity-type systems
to render an application of time-dependent renormalization-group
approaches
\cite{cazalilla2002tDMRGsystmethodquantummanybodooeq,white2004realtimevDMRG,anders2005realtimedynquantimpsys-tDMRG}
possible.  With the present paper we study a simplified problem with a
drastically reduced Hilbert space dimension and focus on two and four
spinful fermions with on-site ($U$) and nearest-neighbor interaction
($V$) on large one-dimensional lattices only.  The time evolution of
this few-fermion quantum system is accessible by a numerically exact
Krylov-space
approach.\cite{lanczos1950iteration,park1986unitary,saad1992analysiskrylovsubspaceapprox,hochbruck1997krylov,hochbruck1999exponentialintegrators,molervanloan2003expmatrix,manmana2005timev1d}
Our study tries to shed some light on the following issues discussed
extensively in the recent literature:

For a single doublon prepared at a definite site initially, we show
how the resulting propagation pattern is affected by $U$ and $V$ and
how this is understood in terms of perturbative arguments in the
strong-coupling limit.  The manifestation of ``energy conservation''
will be analyzed by studying the short-time dynamics of a doublon.

The real-time dynamics of a quantum system in a highly excited state
on the one hand and the spectrum of excitations out of thermal
equilibrium, as obtained in linear-response theory, on the other hand
are usually two completely different issues.  Here, we discuss a
one-to-one relation that is obtained for the case of a single doublon
and therewith address the physics of the long-time stability of a
single doublon.

The real-time dynamics of two doublons in different initial states is
discussed.  Particularly, the $V$ dependencies are interesting as
there is a reduced effective doublon-doublon attraction in the Fermi
opposed to the Bose case which is important to understand the
competition between Bose condensation and phase separation of
doublons.\cite{petrosyan2007quantumliquid,rosch2008metastable}

For a thermodynamically relevant number of fermions, one generally
expects that with the presence of many additional degrees of freedom
there is an enhanced probability for doublon decay since the doublon
energy can be accommodated among different particles in a high-order
scattering event.  This is already seen by means of the ladder
approximation applicable to the low-density limit where a strong
initial decay at short times is observed followed by a slow
exponential decay at long times.\cite{hansen2011splithubbardbands}
Here, this question is studied for the case of four fermions and
discussed in the context of recent time-dependent density-matrix
renormalization-group
studies.\cite{hassanieh2008excitonsin1dhubbard,enss2011lightconerenorm}

The paper is organized as follows: The next section, Sec.\ \ref{sec:theory},
introduces the model, the central observables and the Krylov approach.
We start with the analysis of single-doublon propagation in Sec.\
\ref{sec:prop}, discuss the effects of the nearest-neighbor
interaction in Sec.\ \ref{sec:nnint} and the short-time decay in Sec.\
\ref{sec:decay}.  The relation to APS is worked out in Sec.\
\ref{sec:aps}, and the long-time stability is discussed in Sec.\
\ref{sec:decaylong}.  The second part of the paper is devoted to our
four-fermion results: We discuss the dynamics of two doublons in Sec.\
\ref{sec:four-fermions} and doublon-fermion scattering in Sec.\
\ref{sec:df}.  Final remarks and conclusions are given in Sec.\
\ref{sec:summary}.


\section{Extended Hubbard model and basic theory}
\label{sec:theory}

Ultracold atoms, loaded into an optical lattice, are subject to
different kinds of interaction.
\cite{bloch2008manybodyultracoldgases,trefzger2011ultracolddipgasesoptlatt,lewenstein2007ultracoldatomicgases}
In the simplest cases these are short-ranged, like van der Waals
forces, scaling as $1/r^6$ and hence approximately act on-site only.
Depending on the atomic species, however, more general interactions
can occur.  For example, polarized dipolar atoms experience a
dipole-dipole interaction given by $U_\text{dd} \propto
(1-3\cos^2\theta)/r^3$.  Depending on the angle $\theta$ between the
dipole moments and their relative displacement vector, this can either
be repulsive or attractive.  It is comparatively long-ranged and
usually modeled as an interaction between nearest neighbors.  Overall,
this motivates the extended Hubbard model:
\begin{multline}
  \op H = -\hop \sum_{\left<ij\right>}\sum_{\sigma} \cc{i\sigma}\ca{j\sigma} + U
  \sum_i \nc{i\da}\nc{i\ua} \\ + \frac{V}{2}
  \sum_{\left<ij\right>}\sum_{\sigma\sigma'}\nc{i\sigma}\nc{j\sigma'} =: \Hhop +
  \op H_U + \op H_V \, , \label{eq:hubbard-model}
\end{multline}
which also applies as a model description to electrons interacting via
the screened Coulomb repulsion in condensed-matter systems, e.g.\
transition-metal compounds, if orbital degrees of freedom can be
neglected.  Here, $i$ and $j$ refer to the sites of a one- or
higher-dimensional lattice, $\left<ij\right>$ denotes nearest neighbors, and
$\sigma=\ua,\da$ is the spin projection.  $J$ is the nearest-neighbor
hopping, and $U$ and $V$ the on-site and the nearest-neighbor
interaction strength.  

Our central object of interest is the time-dependent expectation value
of both the local and the total double occupancy, namely $\left< \op
  D_i (t) \right>$ and $\left< \op D (t) \right> = \sum_i \left< \op
  D_i (t) \right>$, respectively.  Here, the local double-occupation
operator is given by $\op D_i = \nc{i\ua}\nc{i\da}$ where
$\nc{i\sigma} = \cc{i\sigma}\ca{i\sigma}$ is the number operator and
where $c^{(\dagger)}_{i\sigma}$ denotes the annihilation (creation) operator for
a fermion at site $i$ with spin $\sigma$.  The time dependence of the
expectation value is due to the time dependence of the system's state
$\ket|\psi(t)> = \exp(-i \op Ht) \ket|\psi_\text{ini}>$ where
$\ket|\psi_\text{ini}>$ is the state in which the system was prepared
initially at time $t=0$.

For systems with moderately large Hilbert-space dimensions $d$, the
numerically exact time evolution of a given initial state is
accessible by means of a time-dependent Krylov-space
technique.
\cite{lanczos1950iteration,park1986unitary,saad1992analysiskrylovsubspaceapprox,hochbruck1997krylov,hochbruck1999exponentialintegrators,molervanloan2003expmatrix,manmana2005timev1d}
Some details of the method are summarized in Appendix \ref{sec:krylov}.

In the following we concentrate on a one-dimensional lattice with $L$
sites and two or four fermions with equal number of up and down spins.
Thereby different processes, such as the propagation and decay of a
single doublon as well as doublon-fermion and doublon-doublon
scattering, can be studied.  For two fermions, the Hilbert-space
dimension is $d = L^2$ and we opt for a lattice with $L=100$ sites.
For four fermions, it is $d = L^2 (L-1)^2 / 4$ and we shorten the
lattice to $50$ sites.  In either case, periodic boundary conditions
are assumed.

\section{Propagation of a single doublon}
\label{sec:prop}

\begin{figure}[!t]
  \includegraphics[width=\columnwidth]{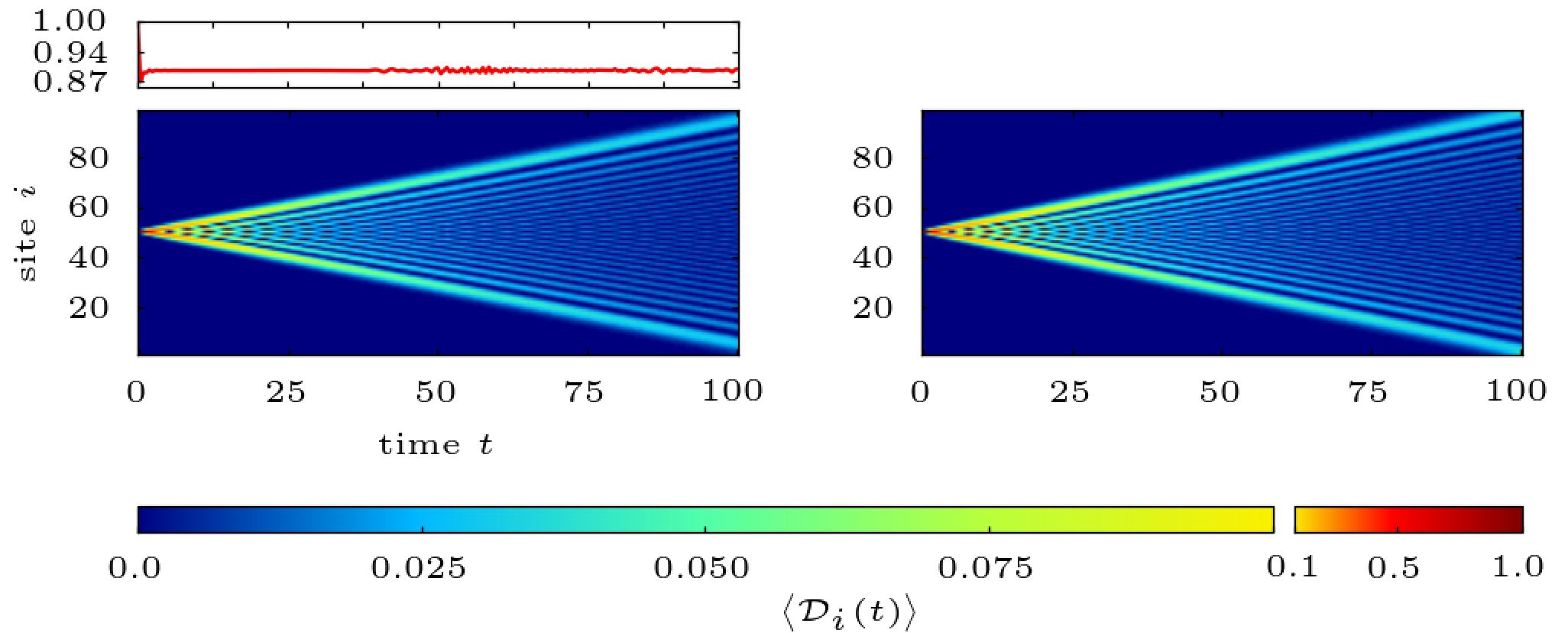}
  \caption{Temporal evolution of the expectation value (see color
    code) of the double occupancy at sites $i=1,...,L$.  {\em Left:}
    Numerical results for the one-dimensional Hubbard model ($V=0$)
    with $L=100$ sites and periodic boundary conditions at
    $U=8\hop$. Initially, at $t=0$, the two-fermion state has been
    prepared as a doublon at $i_{0}=50$.  {\em Top:} Time evolution of
    the expectation value of the total double occupancy.  {\em Right:}
    Corresponding analytical results of the effective model, see Eq.\
    (\ref{eq:loc-docc-bessel}).  Note that most of the color range is
    used to display expectation values less than $0.1$, as shown in
    the color bar below. The time $t$ is measured in units of the
    inverse hopping $1/\hop$.}
  \label{fig:eff}
\end{figure}

To begin with, we consider the two-fermion system and assume that
initially, at time $t=0$, both fermions are at the same site $i_{0}$,
i.e.\ $|\psi_\text{ini} \rangle = \cc{i_{0}\ua}\cc{i_{0}\da} \ket|0>$.
Figure~\ref{fig:eff} (left part) shows the time evolution of the
expectation value of the local double occupancy at $V=0$ and for
strong on-site interaction $U=8J$.  The nearest-neighbor hopping $J=1$
fixes the energy and time scales.  We notice different effects.  First
of all, the doublon delocalizes.  The double occupancy $\left< \op D_i
  (t) \right>$ at the site where the doublon has been prepared
initially ($i_{0}=50$) quickly decreases, and in the course of time
$\left< \op D_i (t) \right>$ basically spreads out over the entire
lattice.  For the time scale $t<100$ shown in the figure, the ``light
cones'' do not yet interfere through the periodic boundary.  Second,
there is doublon decay.  The top panel of Fig.~\ref{fig:eff} shows the
total double occupancy $\left< \op D (t) \right>=\sum_{i}\left< \op
  D_i (t) \right>$.  There is a significant decay from the initial
value $\left< \op D (t) \right>=1$ to about $\left< \op D (t) \right>
\approx 0.9$ in a very short time $t\lesssim 0.5$ (not resolved on the
scale of the figure), followed by an almost constant trend.  The tiny
fluctuations around the constant ``final'' value are simply reflecting
the fact that the total double occupancy does not commute with the
Hamiltonian.

Except for the decay of the doublon, all the details of the entire
propagation profile are fully captured by a simple analytical
description in an effective low-energy model; see the right panel of
Fig.~\ref{fig:eff}.  
As described in Appendix \ref{sec:effective}, this effective model is 
obtained by a unitary transformation
to project out the energetically well separated high-energy part of
the spectrum, thereby generating effective low-energy couplings
perturbatively, in powers of $J/U$:
\cite{foldy1950diractheory,chao1977degperturbationtheory,chao1977kineticexchange,spalek2007tJmodel,fazekas1999lecturenoteselectroncorr}
\begin{equation}
  \label{eq:Heff-docc}
  \op H_{\text{eff}}^{(d)} = \frac{\hop'}{2} \sum_{\left<ij\right>} \cre{d}{i} \ann{d}{j} + ( \hop' + U )
  \sum_i \n{d}{i} - \frac{\hop'}{2} \sum_{\left<ij\right>} \n{d}{i} \n{d}{j} \, . 
\end{equation}
Here, $\cre{d}{i} = \cc{i\ua}\cc{i\da}$ and $\ann{d}{i}$ describe hard-core bosons with
the constraint $\cre{d}{i}{}^{2} = 0$.  Furthermore, $\n{d}{i} =
\cre{d}{i}\ann{d}{i} = \nc{i\ua}\nc{i\da}$ is the local doublon
number.  Hence Eq.\ \eqref{eq:Heff-docc} involves doublon degrees of freedom only and 
takes the form of an extended Bose-Hubbard model with the effective hopping 
$\hop' = 4\hop^2/ U$ and an (in case of positive $U$) attractive
nearest-neighbor interaction.

\begin{figure*}[!t]
  \subfloat[$U=0,V=-10$]{\label{fig:oneU0V-10}\includegraphics[width=.2\textwidth]{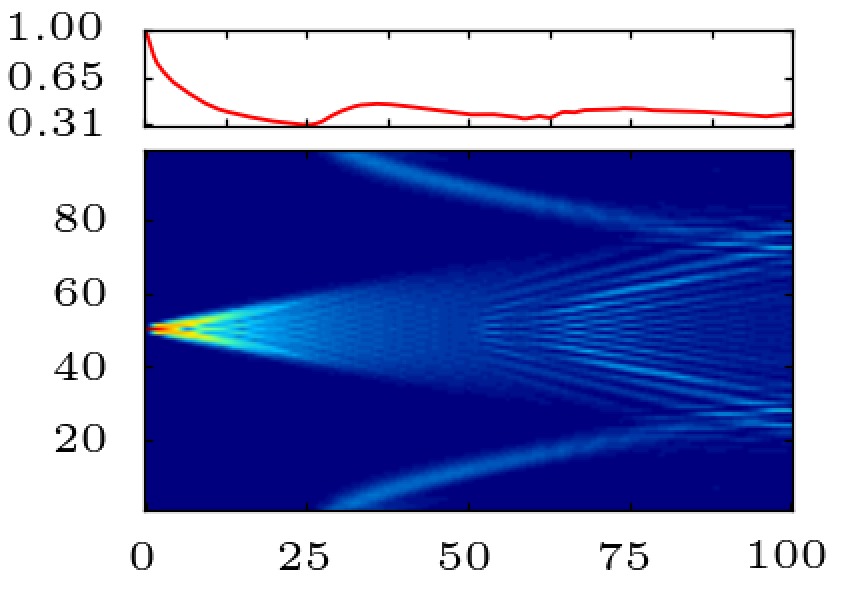}}%
  \subfloat[$U=0,V=-5$]{\label{fig:oneU0V-5}\includegraphics[width=.2\textwidth]{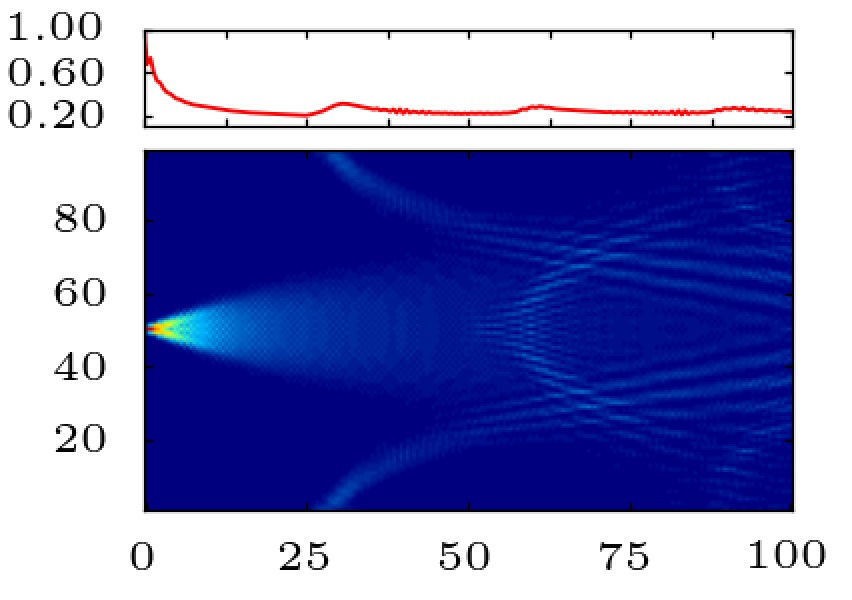}}%
  \subfloat[$U=0,V=0$]{\label{fig:oneU0V0}\includegraphics[width=.2\textwidth]{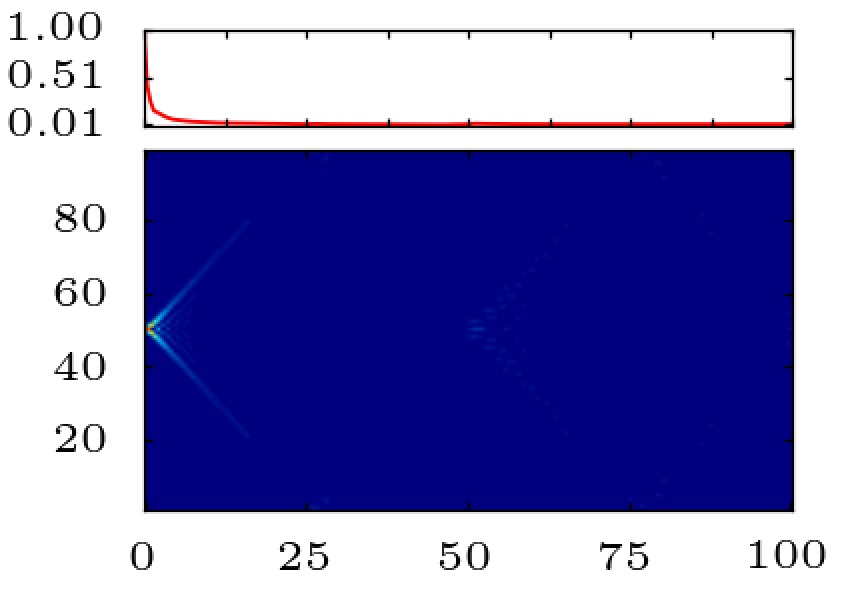}}%
  \subfloat[$U=0,V=5$]{\label{fig:oneU0V5}\includegraphics[width=.2\textwidth]{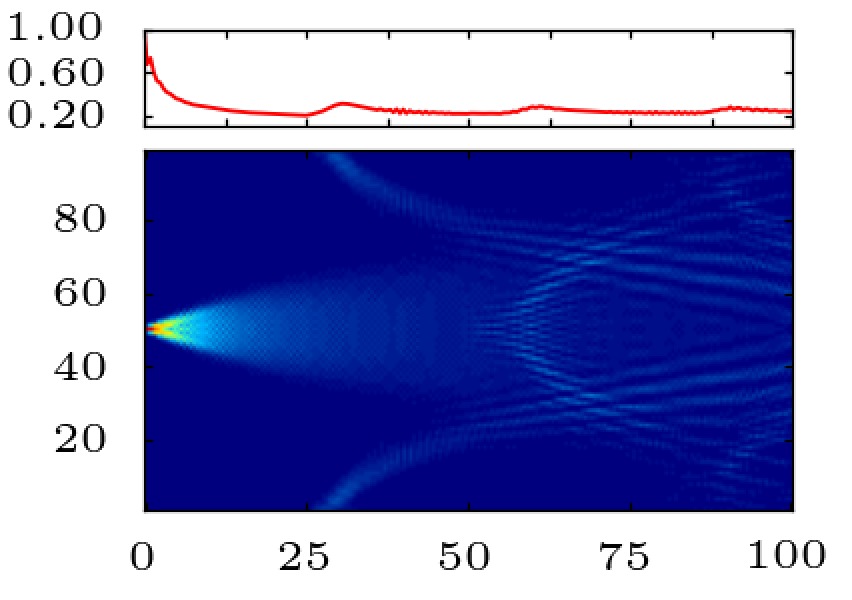}}%
  \subfloat[$U=0,V=10$]{\label{fig:oneU0V10}\includegraphics[width=.2\textwidth]{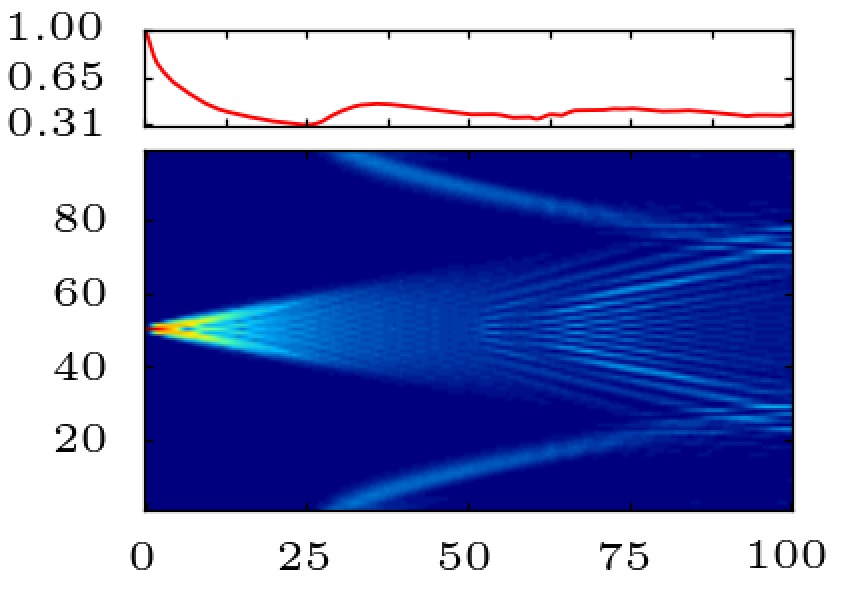}}%
  \\
  \subfloat[$U=5,V=-10$]{\label{fig:oneU5V-10}\includegraphics[width=.2\textwidth]{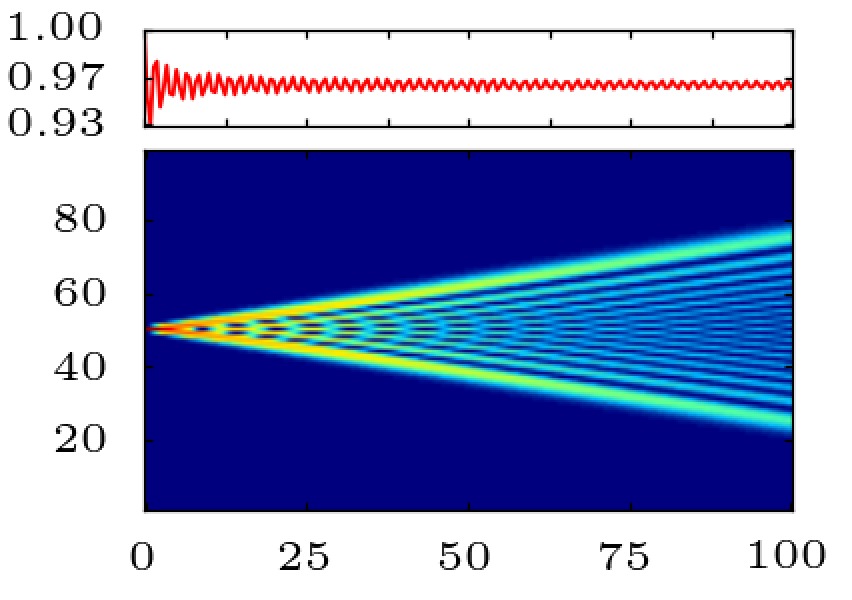}}%
  \subfloat[$U=5,V=-5$]{\label{fig:oneU5V-5}\includegraphics[width=.2\textwidth]{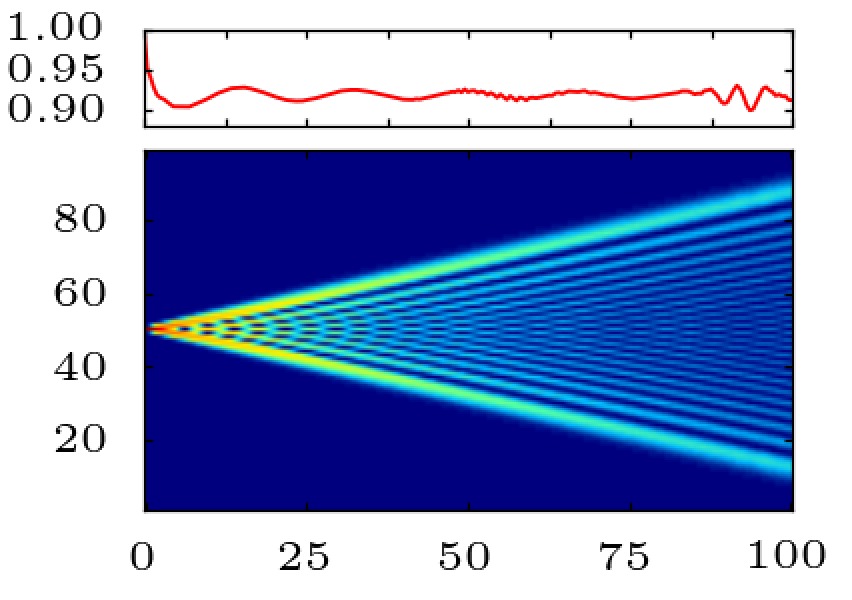}}%
  \subfloat[$U=5,V=0$]{\label{fig:oneU5V0}\includegraphics[width=.2\textwidth]{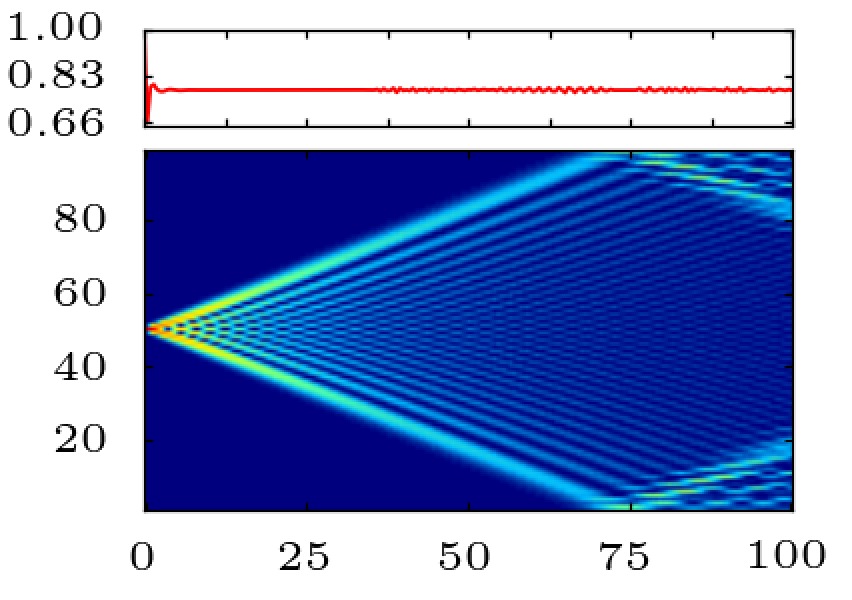}}%
  \subfloat[$U=5,V=5$]{\label{fig:oneU5V5}\includegraphics[width=.2\textwidth]{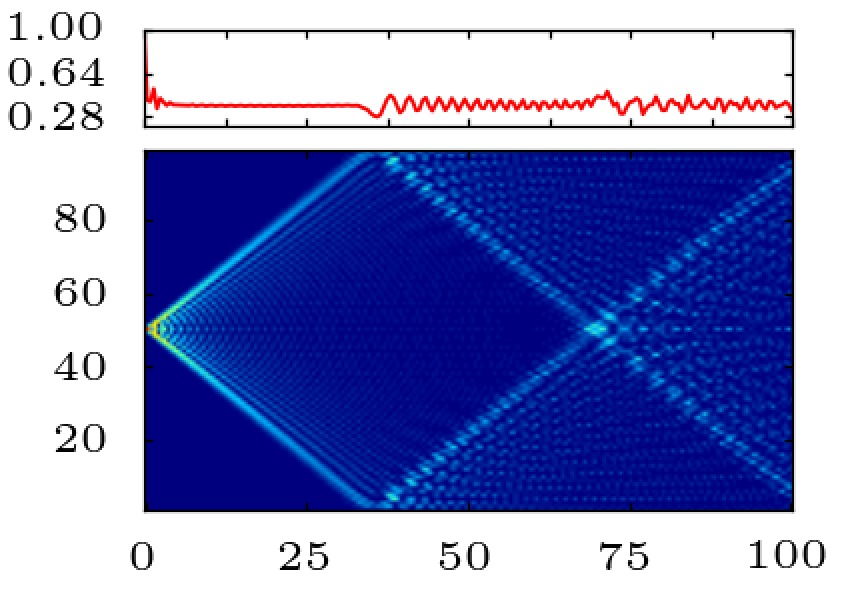}}%
  \subfloat[$U=5,V=10$]{\label{fig:oneU5V10}\includegraphics[width=.2\textwidth]{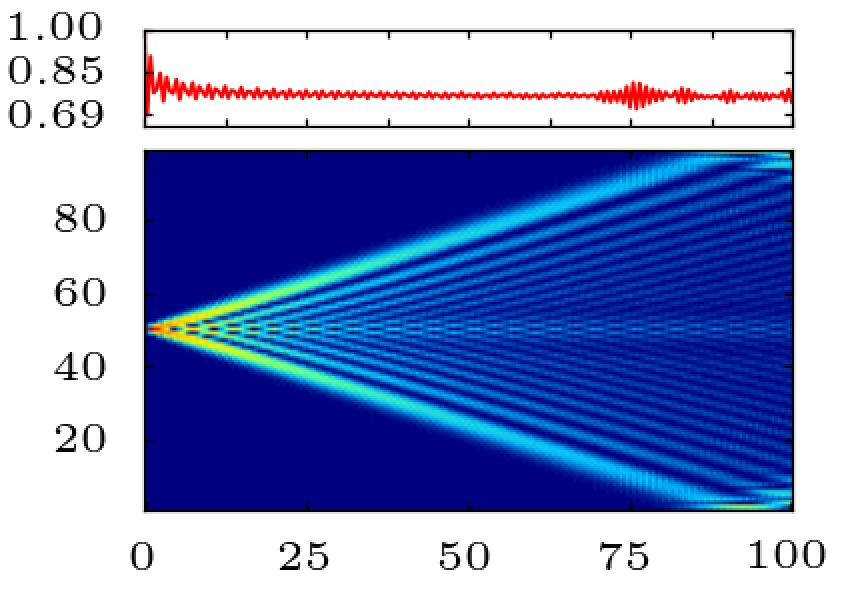}}%
  \\
  \subfloat[$U=10,V=-10$]{\label{fig:oneU10V-10}\includegraphics[width=.2\textwidth]{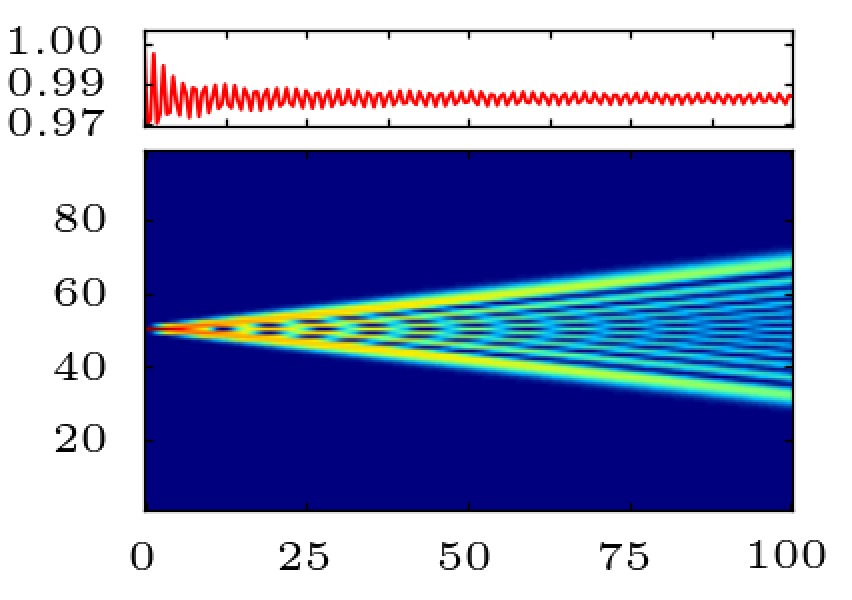}}%
  \subfloat[$U=10,V=-5$]{\label{fig:oneU10V-5}\includegraphics[width=.2\textwidth]{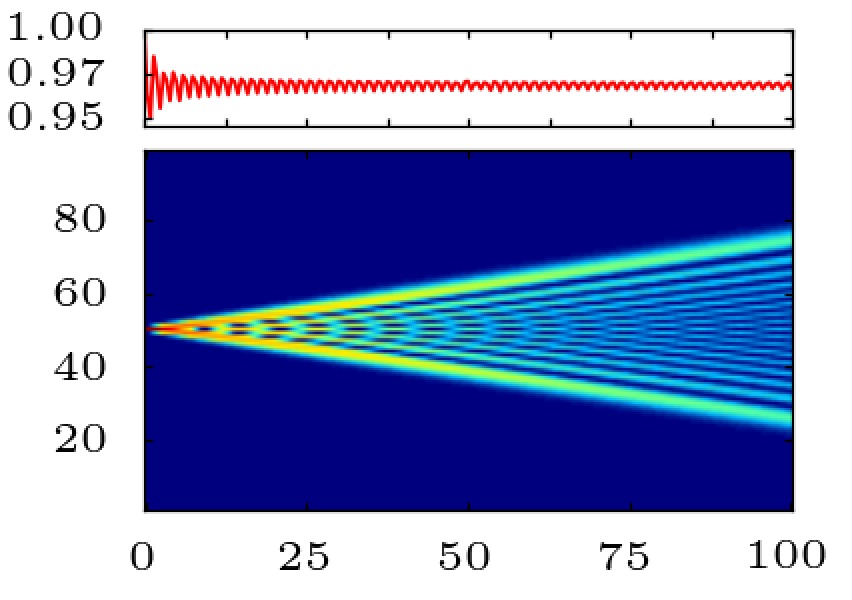}}%
  \subfloat[$U=10,V=0$]{\label{fig:oneU10V0}\includegraphics[width=.2\textwidth]{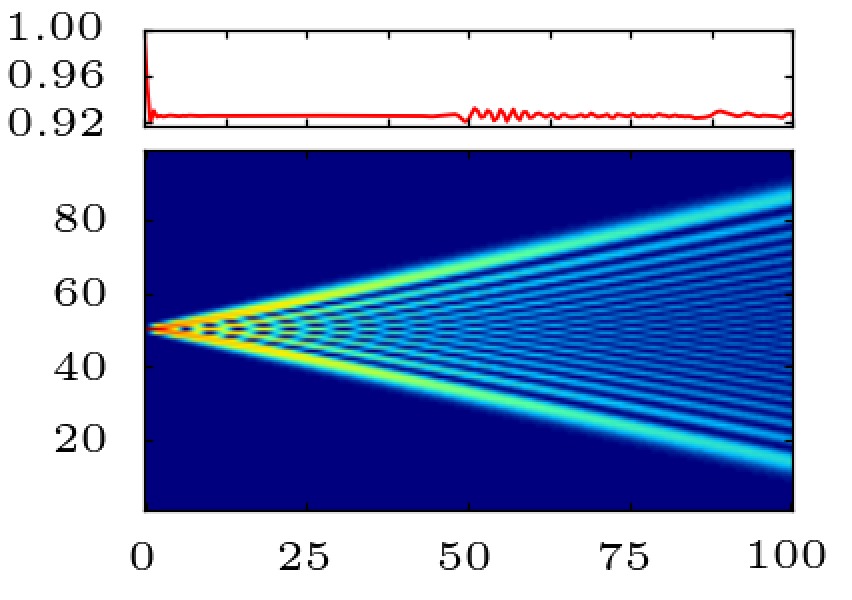}}%
  \subfloat[$U=10,V=5$]{\label{fig:oneU10V5}\includegraphics[width=.2\textwidth]{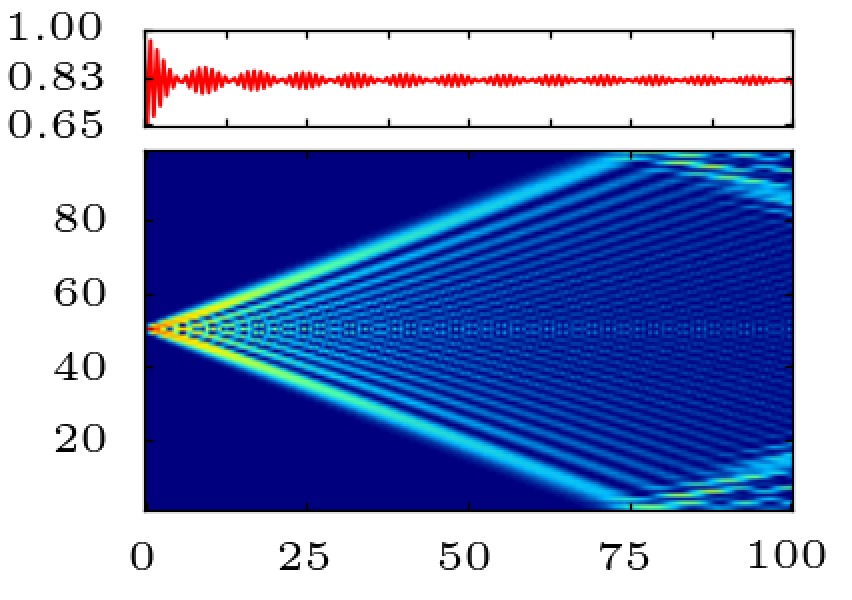}}%
  \subfloat[$U=10,V=10$]{\label{fig:oneU10V10}\includegraphics[width=.2\textwidth]{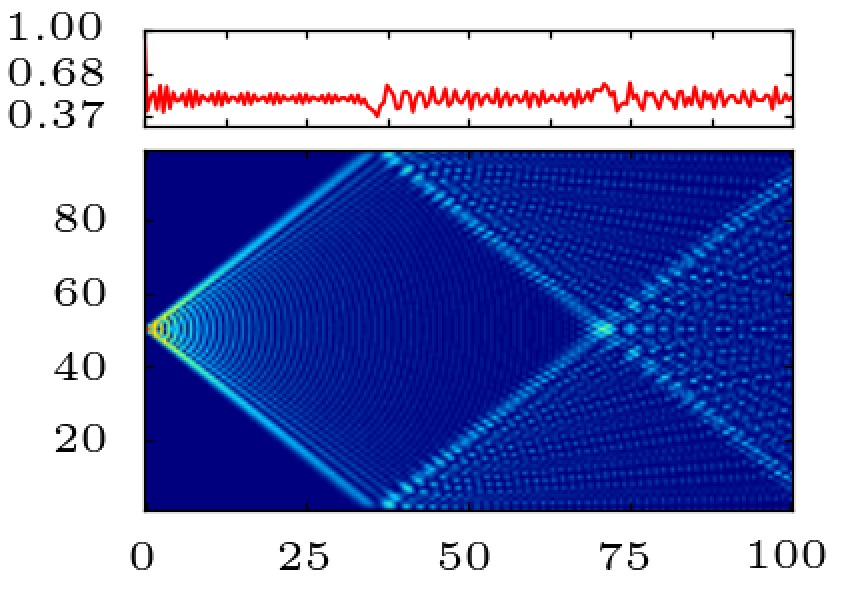}}%
  \caption{ Time-dependent expectation value of the local (main
    panels) and total double occupancy (small top panels) for two
    fermions initially prepared at the same site $i_{0}$ of a
    one-dimensional lattice with $L=100$ sites and periodic boundary
    conditions.  Results for on-site interaction $U = 0,5,10$ and
    nearest-neighbor coupling $V=0,\pm 5,\pm 10$, as indicated.  Note
    that for the total double occupancy the scale of the $y$-axis
    differs from case to case.  The color code is the same as in Fig.\
    \ref{fig:eff} and the same for all plots.}
  \label{fig:one}
\end{figure*}

For a system with a single doublon only, the interaction term can be
disregarded, and the resulting free tight-binding Hamiltonian is
diagonalized by Fourier transformation.  In the limit
$L\rightarrow\infty$, the time-dependent local double occupancy in the
effective model is then found to be given by the $k$th Bessel
function of the first kind $\mathcal J_k$,
\begin{equation}
  \left< \op D_i^{\text{eff}}(t) \right> = \mathcal J_{i-i_{0}}^2(J't) \, ,
  \label{eq:loc-docc-bessel}
\end{equation}
if the doublon was prepared at site $i_{0}$ initially.  Note that the
total double occupancy is conserved, since $\sum_{k=-\infty}^{\infty}
\mathcal J_k^2(x) = 1$ for all $x$.

The time dependence of the expectation value of the local double
occupancy, as given by Eq.\ \eqref{eq:loc-docc-bessel}, is shown in
Fig.~\ref{fig:eff} (right).  While effects due to doublon decay are
neglected at this level, doublon-propagation effects should be
captured qualitatively correct.  Comparing with the exact numerical
result (Fig.~\ref{fig:eff}, left), we note that the effective model
provides an excellent description of the propagation already for
$U=8J$.

The effect of varying $U$ can be seen in Fig.\ \ref{fig:one}.  The
panels Fig.\ \subref*{fig:oneU0V0}, \subref*{fig:oneU5V0}, and
\subref*{fig:oneU10V0} give the result of the full model for $U=0$,
$U=5J$ and $U=10J$.  We note that the mobility of the doublon
decreases with increasing $U$ which, in the effective model, is due to
the reduced doublon hopping $\sim 1/U$.  The interference pattern
visible for $U=5J$ in panel Fig.\ \subref*{fig:oneU5V0} is due to the
finite system size and periodic boundary conditions.  Apart from that,
however, the pattern does not change much qualitatively as compared to
$U=10J$.  This is worth mentioning since $U=5J$ is well below the
critical $U$ (of the order of twice the free bandwidth $2W=8J$) at
which the two-particle excitation spectrum, related to the APS Green's
function
$\green{\ca{i\sigma}\ca{i\bar\sigma}}{\cc{i\sigma}\cc{i\bar\sigma}}$,
does change qualitatively since the correlation satellite splits off
(see Ref.\ \onlinecite{nolting1990influence}, for example).  This
reminds us that there is a clear conceptual difference between the
two-particle spectrum that refers to excitations starting from the
system's ground state on the one hand and the temporal evolution of a
highly excited initial state on the other.

\section{Effects of nearest-neighbor interaction}
\label{sec:nnint}

The remaining panels of Fig.~\ref{fig:one} show propagations patterns
for finite nearest-neighbor interaction $V$.  For $U=10J$, see last
row in Fig.~\ref{fig:one}, we find a decreasing mobility of the
doublon with increasing difference between the on-site and the
nearest-neighbor interaction strengths $U-V$.  Similar to the
discussion in the preceding section, this trend is easily explained in
an effective model that preserves the total double occupancy.  This
can be derived, for example, by standard second-order perturbation
theory around the $J=0$ limit and yields an effective doublon hopping
amplitude
\begin{equation}
  J' = 4 \frac{J^{2}}{U-V} \,.
\end{equation}
This corresponds to a sequence of two virtual hopping processes: In
the first, one of the two fermions composing the doublon hops to a
nearest-neighbor site.  Thereby, for $U>V$ ($U<V$), the energy $U-V$
is gained (has to be paid).  The second nearest-neighbor hopping
process leads to the recombination of the doublon, either at the same
or at one of the adjacent sites.

Looking at the cone angle of the ``light cone'' in the propagation
patterns in the last row and comparing the results with $U-V=20$ to
$U-V=5$, the effective description yields the correct trend: As the
expectation value of the double occupancy in the effective model
depends on the product of $J'$ and $t$ only, see Eq.\
(\ref{eq:loc-docc-bessel}), the time axis scales linearly with
$J'$.  The effective doublon hopping also explains that the patterns
in panels Figs.\ \subref*{fig:oneU5V-10} and \subref*{fig:oneU10V-5} and
the patterns in Figs.\ \subref*{fig:oneU5V-5} and \subref*{fig:oneU10V0} as
well as Figs.\ \subref*{fig:oneU5V0} and \subref*{fig:oneU10V5} are almost
equal as $U-V$ is constant, respectively.

For $U=5J$, see middle row in Fig.~\ref{fig:one}, the results for
$U-V=5$ and $U-V=-5$ differ significantly although they should be
described by the same effective hopping $J'$, apart from the sign.
The sign, however, has no effect.  The difference is rather due to the
residual influence of virtual processes of fourth order in $J$ where
one of the fermions hops two sites away, followed by a recombination
of the doublon.  This leads to an asymmetry between the two cases,
$\pm |U-V|$, since for $U>V$ all three intermediate states have lower
energy while for $U<V$ two states are higher in energy and one lower.
With increasing interaction strengths $U$ and $V$, we find this
asymmetry to be less and less efficient as expected.

As can be seen by comparing Figs.\ \subref*{fig:oneU10V0} and
\subref*{fig:oneU10V5}, for example, the ``speed'' of the doublon on
the light cone increases somewhat less than a factor $2$ although $J'$
is exactly twice as large.  Looking at Eq.\
(\ref{eq:loc-docc-bessel}), this hints to a breakdown of the effective
model with $U-J \to 0$.  In fact, for $U=V$, degenerate perturbation
theory in $J$ must be considered.  Since the states with two fermions
at the same and at neighboring sites have the same unperturbed energy
$U=V$, decay and recombination of the doublon becomes a very efficient
process.  This leads to a maximum mobility as can be seen in
Figs.\ \subref*{fig:oneU5V5} and \subref*{fig:oneU10V10}.

For $U=V$, first-order perturbation theory in $J$ partially lifts the
degeneracy.  Therefore, the resulting effective model actually
describes the motion of a new eigenmode which is a linear combination
of a doubly occupied site with states where the two fermions are found
at adjacent sites.  Rather than doublon propagation, the physically
adequate picture is given by propagation of this extended object which
we will refer to as an ``extended doublon'' in the following.

A description by means of an effective model that preserves the total
double occupancy must break down for $U=0$.  This explains the
qualitatively different propagation patterns in the first row of Fig.\
\ref{fig:one}.  For $V=0$ the pattern is given by $\left< \op
  D_{i}^\text{free} (t) \right> = \mathcal J_{i-i_{0}}^4(2Jt)$.  For
finite $V$, see Fig.\ \subref*{fig:oneU0V10}, for example, we note that
besides the usual propagation pattern describing the delocalization of
the doublon initially prepared at $i_{0}=50$, there is a finite
probability to find a doubly occupied site around $i=1$ at $t\approx
30 J^{-1}$.  The structure further evolves in time and interferes with
the main structure.  This must be considered as a finite-size effect
resulting for $U=0$ from the very fast decay of the doublon into two
independently moving fermions.  Due the periodic boundary condition,
this implies that the two fermions meet again and form a doubly
occupied site.  The corresponding probability strongly decreases with
the system size $L$.

\section{Decay of a doublon at short times}
\label{sec:decay}

The main idea behind the concept of a repulsively bound pair of
fermions is that energy conservation prevents the decay of a doublon
at strong coupling $U>2W$: The doublon energy of the order of $U$
cannot be transferred to two independently moving fermions with a
kinetic energy of the order of at most $W$ each.  In Fig.\
\ref{fig:one}, the small top panels show the time dependence of the
total double occupancy.  In all cases we find a relaxation of the
total double occupancy from its initial unit value to a nearly
constant value after a short time.  In many cases, this quick initial
decay is hardly resolved on the scale of the figure; see $U=5$ and
$U=10$ ($V=0$) for example.

\begin{figure}[t!]
  \includegraphics[width=.8\columnwidth]{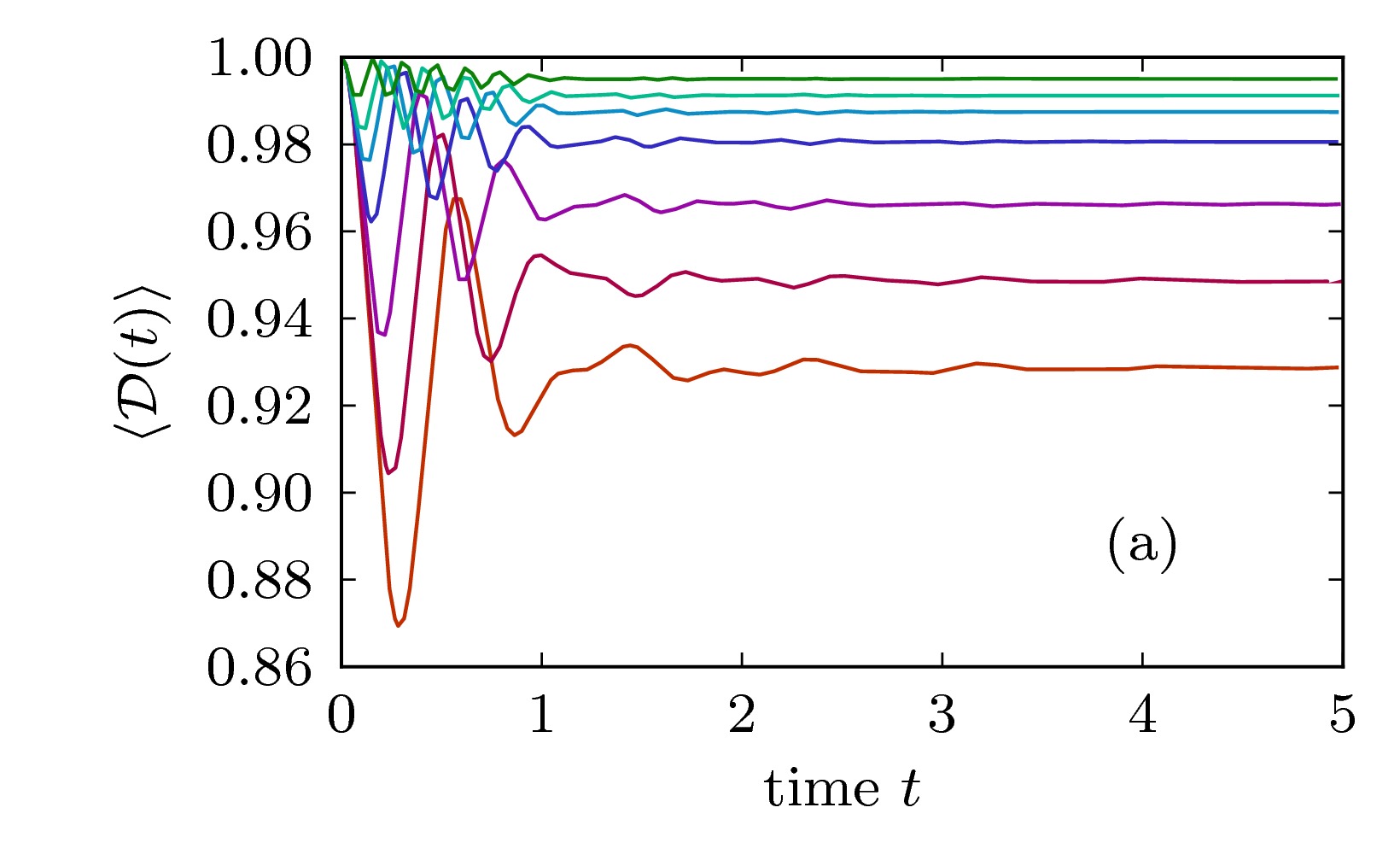}%
  \\
  \includegraphics[width=.8\columnwidth]{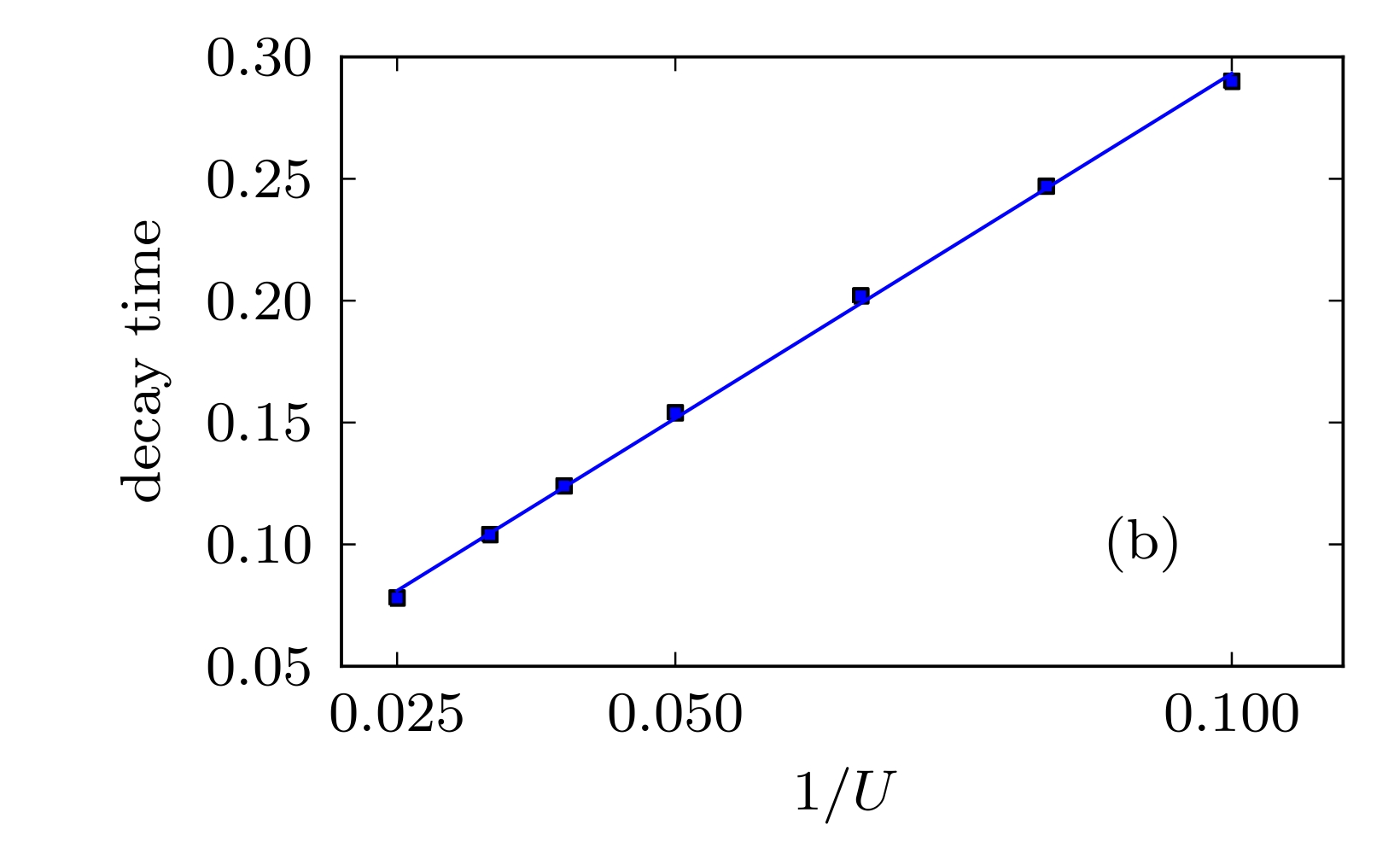}%
  \caption{(a) Short-time behavior of the total double occupancy for
    $U=10,12,15,20,25,30,40$ (from bottom to top) and $V=0$.  (b)
    First local minimum point (``decay time'') of the time-dependent
    total double occupancy plotted against $1/U$.  The line is a guide
    to the eye.  }
  \label{fig:one_shorttdecay}
\end{figure}

Figure\ \ref{fig:one_shorttdecay}(a) shows $\langle \op D (t) \rangle$
for $V=0$ and different $U$ on a much shorter time scale up to a few
inverse hoppings $1/J$.  To quantify the time scale for the doublon
decay, we look at the position of the first minimum.  This ``decay
time'' is shown in Fig.\ \ref{fig:one_shorttdecay}(b) as a function of
$1/U$.  We find a simple linear relation.  The depth of the first
minimum also increases with increasing interaction strength.  In all
cases, however, the double occupancy does not recover completely to
its initial value but after some oscillations relaxes to a nearly
constant value which is becomes smaller for weaker $U$.

The question how the observed doublon decay is consistent with energy
conservation, is easily answered by means of time-dependent
perturbation theory in $J$.  For $J=0$, the total double occupancy is
conserved.  This already explains the high and nearly constant
$\langle \op D (t) \rangle$ for very strong $U$ [see the result for
$U=40$ in Fig.\ \ref{fig:one_shorttdecay}(a)].  For strong but finite
$U$ first-order-in-$J$ time-dependent perturbation theory predicts the
transition probability between two unperturbed energy eigenstates
$|\psi_m \rangle$ and $|\psi_n\rangle$ to behave as
\cite{sakurai1993modernqm}
\begin{equation}
  \label{eq:fopt}
  \left| \langle \psi_n | e^{-i\op H t} \psi_m \rangle \right|^2 \propto \frac{ \sin^2\left( \frac{\Delta E_{m \ra n}}{2} t \right) }{
    (\Delta E_{m \ra n})^2 } \, .
\end{equation}
This reminds us that ``energy conservation'' as used in the argument
given at the beginning of the section holds in the long-time limit
only where the right-hand side of Eq.~\eqref{eq:fopt} evolves into a
$\delta$ function.

Doublon decay is possible (i) at short times or (ii) at long times and
consistent with energy conservation in the presence of additional
degrees of freedom to dissipate the excess energy.  Let us discuss the
case (i) first [see Sec.\ \ref{sec:df} for point (ii)]: As a function
of the energy difference $\Delta E_{m \ra n}$, the transition
probability has a peak structure with a width that scales as $1/t$.
Hence transitions are possible between states with energy difference
$\Delta E_{m \ra n} \lesssim 1/ t$.  To put it in other words,
excitations with energy $\Delta E_{m \ra n}$ most probably occur on a
time scale $t \lesssim 1/\Delta E_{i\ra j}$.  Therefore, since the
dissociation of two fermions in the strong-coupling regime involves
energies of the order of $U$, the position of the first minimum must
scale with $1/U$, as demonstrated in
Fig.~\ref{fig:one_shorttdecay}(b).

At very short times, the decay is independent of the coupling $U$, as
seen in Fig.~\ref{fig:one_shorttdecay}(a) for $t\lesssim 0.2$.  This
is easily explained by Taylor expansion in $t$:
\begin{equation}
  \langle \op D(t) \rangle 
  =
  1 - t^{2} \: \Delta E_{\rm ini}  + {\cal O}(t^{4}) \; ,
\end{equation}
where the variance of the total energy in the initial state is
proportional to the number of nearest neighbors $z=2$,
\begin{equation}
  \Delta E_{\rm ini} = 
  \langle \psi_\text{ini} | \op H^{2} | \psi_\text{ini} \rangle -
  \langle \psi_\text{ini} | \op H | \psi_\text{ini} \rangle^{2}
  =
  2 z J^{2} \; ,
\end{equation}
and thus depends on the hopping amplitude $J$ only.

\section{Doublon dynamics and appearance-potential spectroscopy}
\label{sec:aps}

The time-dependent expectation value of the double occupancy at site
$i$ is
\begin{equation}
  \langle \op D_i(t) \rangle 
  = 
  \langle 0 | \ann{d}{i_{0}} e^{i\op H t} \cre{d}{i} \ann{d}{i} e^{-i\op H t} \cre{d}{i_0} | 0 \rangle
  =
  \big| \langle 0 | \ann{d}{i} e^{-i\op H t} \cre{d}{i_0} | 0 \rangle
  \big|^{2} \,,
  \label{eq:key}
\end{equation}
if a doublon has been prepared at $t=0$ at the site $i_{0}$, i.e.,
$\cre{d}{i} |0\rangle = \cc{i\ua}\cc{i\da} |0\rangle$.  Note that the
original expectation value can be written as a square since (i) $\op
H$ commutes with the total particle number and (ii) we start from the
Fermi vacuum.  Namely, starting from the vacuum state $| 0 \rangle$,
preparing of the doublon at site $i_{0}$, time propagation and finally
annihilation at $i$, we must return to the same state $| 0 \rangle$.

The Fermi vacuum corresponds to an empty band in the context of
electron spectroscopy.  Let us discuss the relation of doublon
dynamics to appearance-potential spectroscopy
(APS),\cite{park1974softxrayaps,ertl1993spinressoftxrayaps,rangelov2000surfacemagneticprop}
in particular.  Consider the following retarded two-particle
(two-electron) Green's function:
\begin{equation}
  G_{ii,i_{0}i_{0}}(t) = - i \Theta(t) \langle 0 | \ann{d}{i} e^{-i\op H t} \cre{d}{i_{0}}| 0 \rangle \: .
\end{equation}
This is a ground-state quantity, the Fourier transform of which,
$G_{ii,jj}(\omega+i0^{+}) = \green{\ca{i\ua} \ca{i_{\da}}}{\cc{j\da}
  \cc{j\ua}}_\omega$, yields the appearance-potential spectrum
$A_{ii,ii}(\omega) = - \mbox{Im}
G_{ii,ii}(\omega+i0^{+})/\pi$.\cite{potth2001theor}
$A_{ii,ii}(\omega)$ describes the cross section in a non-radiative
two-electron process where an initial electron at high kinetic energy
occupies an empty state in the valence band of a metal by transferring
the energy difference to a core electron which is lifted to another
empty state in the band.  The process is essentially local and
represents the ``time inverse'' of high-resolution
core-valence-valence (CVV) Auger-electron
spectroscopy.

For an empty band, the equation of motion for the APS Green's function
is readily solved:\cite{nolting1990influence}
\begin{equation}
  G_{ii,jj}(\omega+i0^{+}) 
  =
  \frac{1}{L} \sum_{\ff k} e^{i\ff k(\ff R_{i} - \ff R_{j})}
  \frac{\Lambda_{\ff k}(\omega+i0^{+})}{1-U\Lambda_{\ff
      k}(\omega+i0^{+})} \,,
  \label{eq:eom1}
\end{equation}
with
\begin{equation}
  \Lambda_{\ff k}(\omega)
  =
  \frac{1}{L} \sum_{\ff p} \frac{1}{\omega - \varepsilon(\ff p) - \varepsilon(\ff k - \ff p)} \; .
\end{equation}
Here $\ff R_{i}$ denotes the position vector to the site $i$, $\ff k$
is a wave vector of the first Brillouin zone, and the dispersion of
the tight-binding band $\varepsilon(\ff k) = -J \sum_{\ff \Delta}
\exp(-i\ff k \ff \Delta)$ is obtained as a sum over nearest-neighbors
displacement vectors $\ff \Delta$.

For $t>0$ we have $\langle \op D_i(t) \rangle =
|G_{ii,i_{0}i_{0}}(t)|^{2}$ and thus
\begin{equation}
  \langle \op D_i(t) \rangle 
  =
  \left|
    \frac{1}{2\pi} \int d\omega\: e^{-i\omega t} G_{ii,i_{0}i_{0}}(\omega+i0^{+}) 
  \right|^{2} \: .
  \label{eq:eom2}
\end{equation}
At $U=0$ this related to the Bessel function, $\left< \op
  D_{i}^\text{free} (t) \right> = \mathcal J_{i-i_{0}}^4(2Jt)$.  For
$U>0$, and using the fact that the Green's function is the Hilbert
transform of the spectral function, we find
\begin{equation}
  \langle \op D_i(t) \rangle 
  =
  \int d\omega \int d\omega' e^{i(\omega - \omega')t} A_{i_{0}i_{0},ii}(\omega) A_{ii,i_{0}i_{0}}(\omega') \: .
  \label{eq:doft}
\end{equation}
After substituting $\omega \mapsto \omega + \omega'$, we see that the
time dependence of the local double occupancy is given by the Fourier
transform from frequency to time representation of the
self-convolution of the APS spectral function.  This relation is
remarkable as it provides a link between the APS spectral function, an
equilibrium quantity describing two-particle excitations within the
framework of linear-response theory, and the non-equilibrium time
evolution of the local double occupancy.  It is by no means general,
however, and can be traced back to Eq.\ (\ref{eq:key}) which holds in
the case of an empty band only.

\section{Decay of a doublon --- long-time stability}
\label{sec:decaylong}

\begin{figure}[!t]
  \includegraphics[width=\columnwidth]{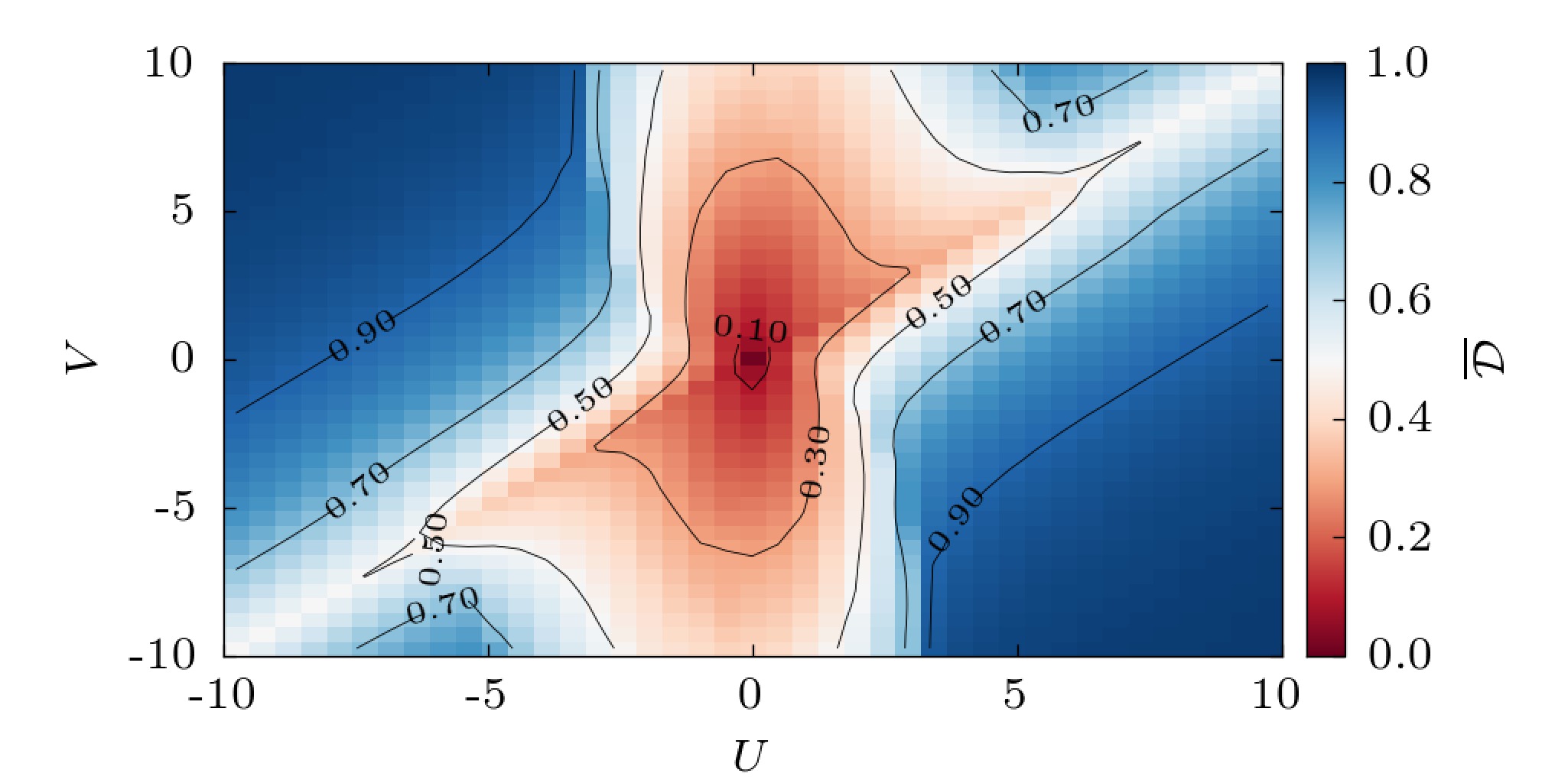}%
  \\
  \includegraphics[width=\columnwidth]{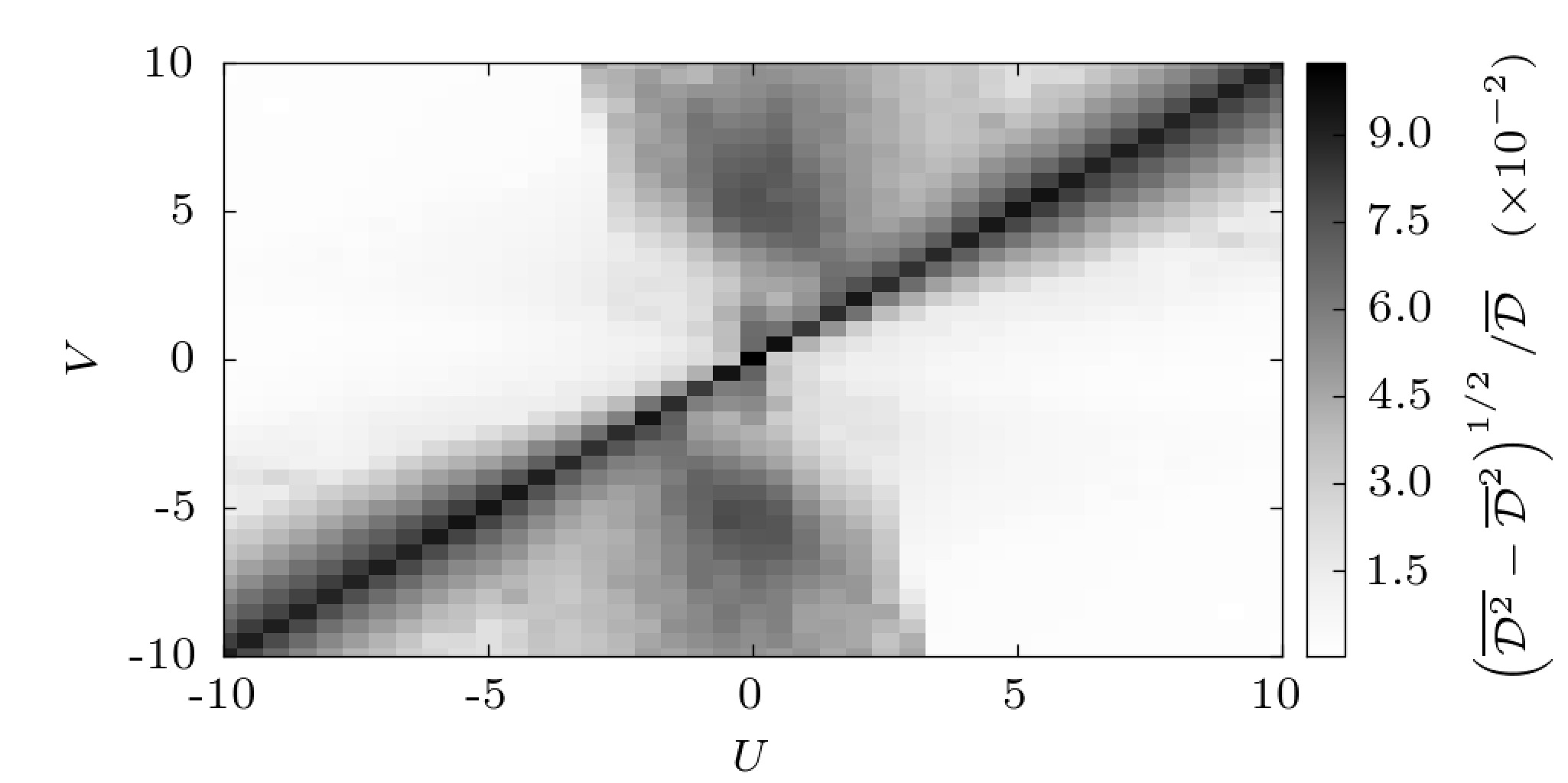}%
  \caption{ Long-time average (top) and relative standard deviation
    (bottom) of the total double occupancy $\langle \op D (t) \rangle$
    for interaction strengths $-10<U<10$ and $-10<V<10$.
    $\overline{\op D}$ and $(\overline{\op D^{2}} - \overline{\op
      D}^{2})^{1/2}/\overline{\op D}$ are calculated for the time
    interval $50<t<100$.  The color or grey-scale code is given on the
    right.  }
  \label{fig:one_stab}
\end{figure}

\begin{figure}[!t]
  \centering
  \includegraphics[width=.5\columnwidth]{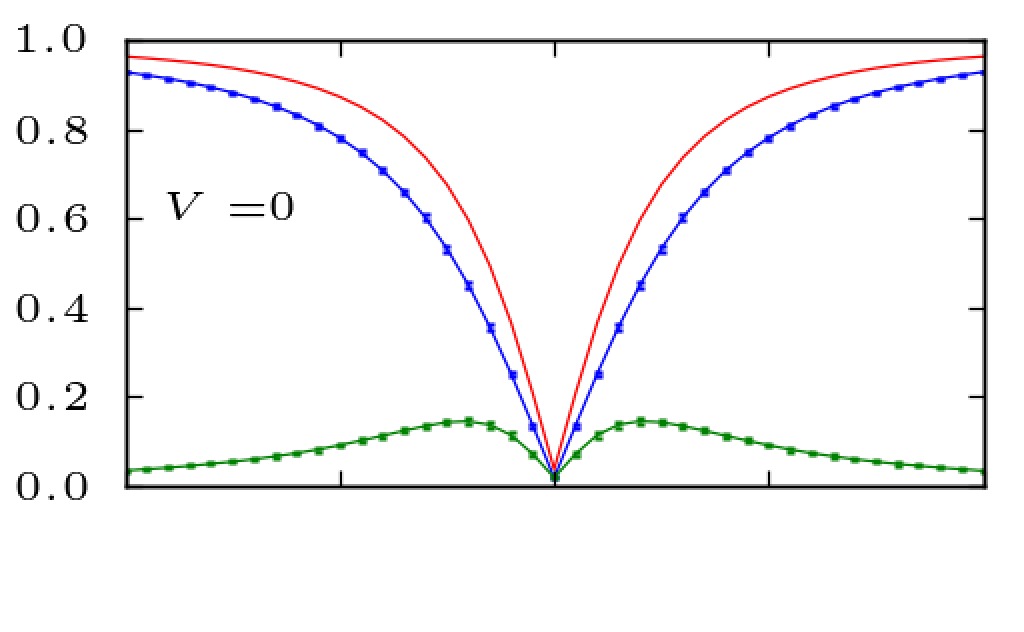}%
  \includegraphics[width=.5\columnwidth]{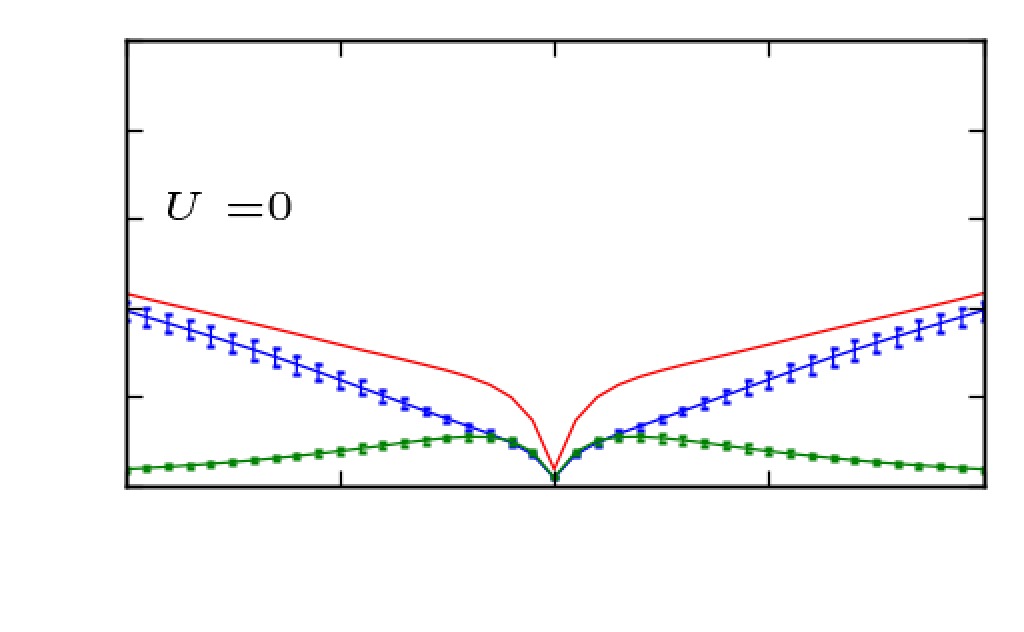}%
  \\
  \includegraphics[width=.5\columnwidth]{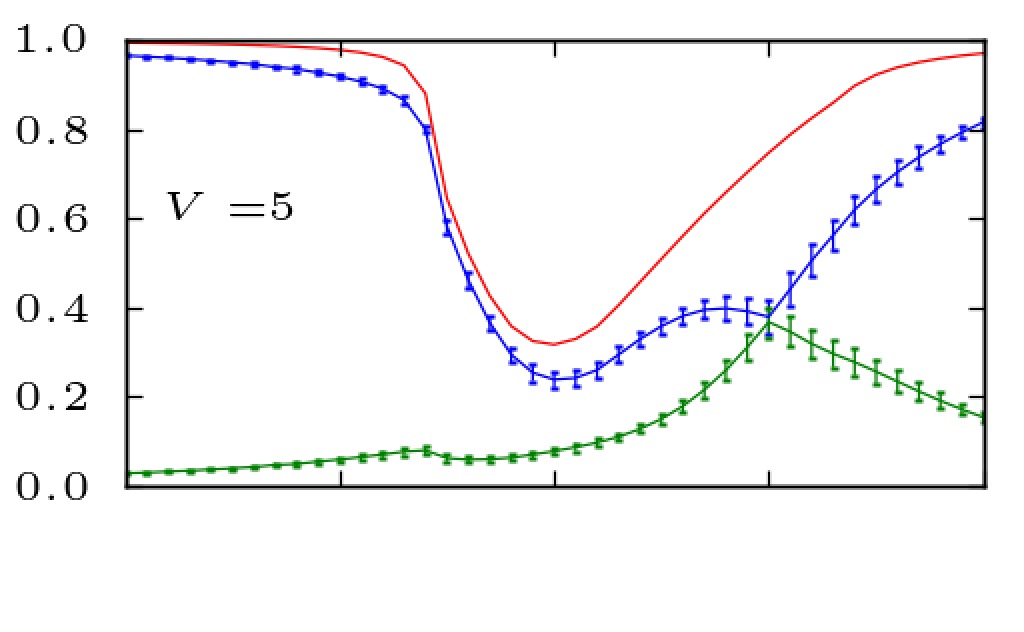}%
  \includegraphics[width=.5\columnwidth]{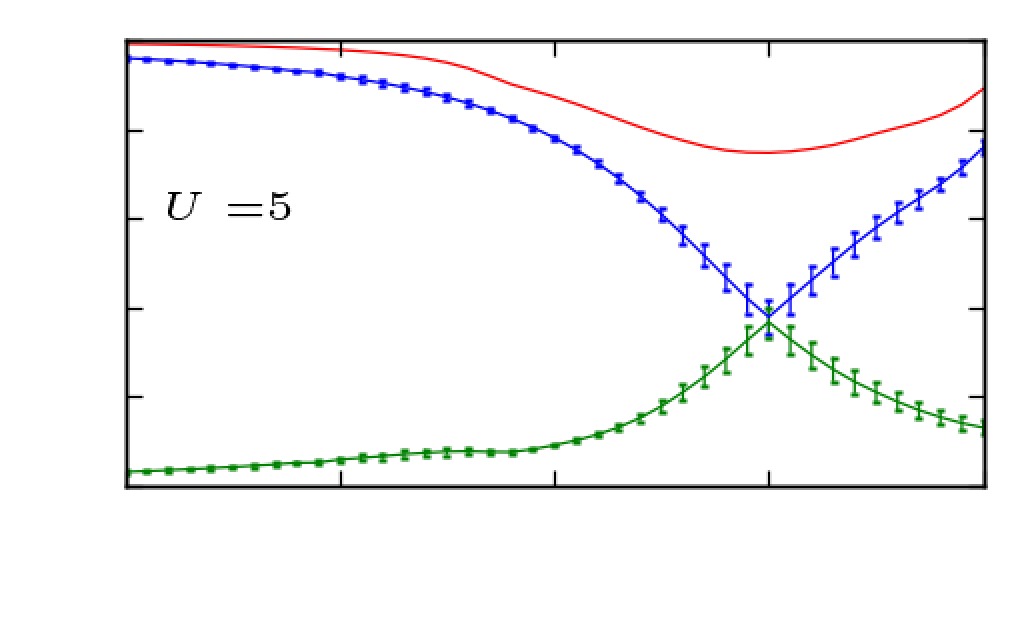}%
  \\
  \includegraphics[width=.5\columnwidth]{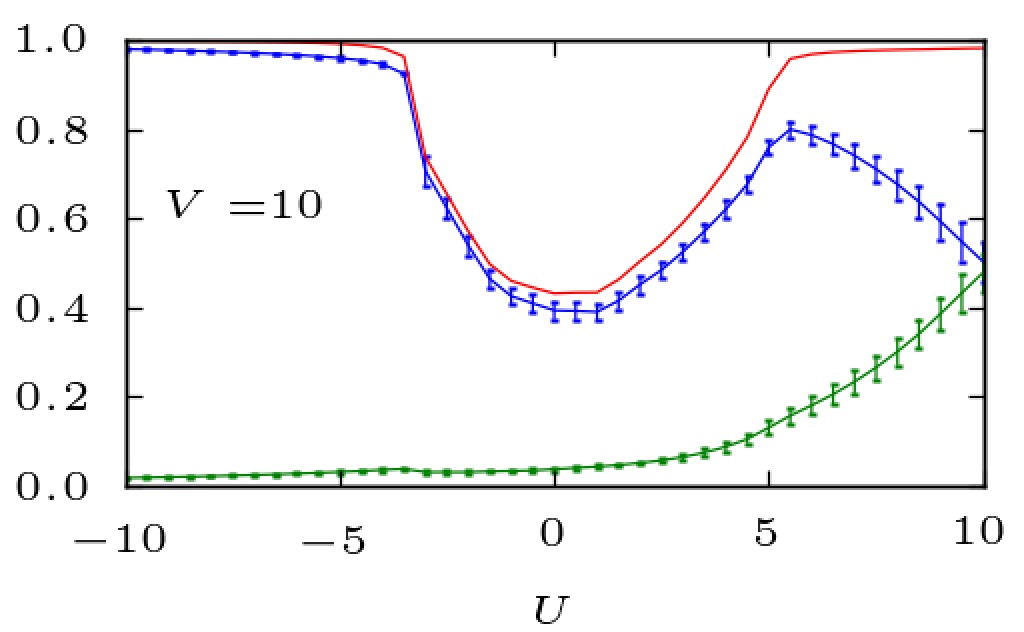}%
  \includegraphics[width=.5\columnwidth]{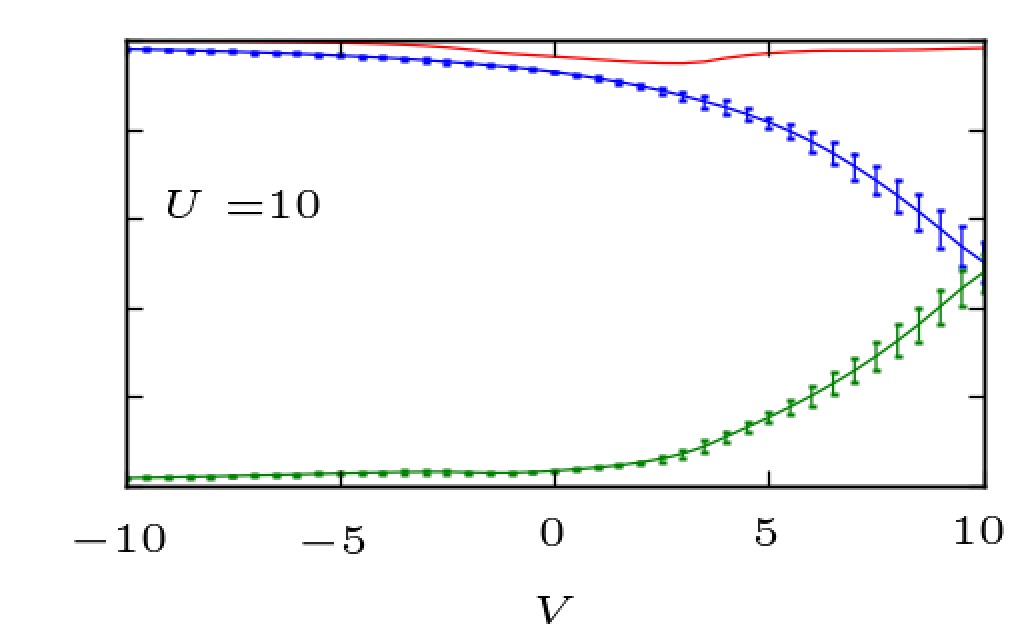}%
  \caption{ Sectional views of the stability map
    Fig.~\ref{fig:one_stab} for $U,V=0,5,10$.  Blue lines show
    $\overline{\op D}$, error bars indicate the absolute standard
    deviation.  Green lines with error bars: time average and standard
    deviation of the nearest-neighbor occupancy, $\langle \frac{1}{2}
    \sum_{<ij>} \sum_{\sigma\sigma'} \nc{i\sigma}(t) \nc{j\sigma'}(t)
    \rangle$.  Red: sum of double and nearest-neighbor occupancy.  }
  \label{fig:sec_one_stab}
\end{figure}

In the long-time limit, for an infinitely large system, i.e.,
$L\to\infty$, the local double occupancy $\langle \op D_{i} (t)
\rangle \to 0$ for $t\to\infty$ due to a complete delocalization of
the doublon or the two independent fermions, respectively.  The total
double occupancy $\langle \op D (t) \rangle$, however, may relax to a
finite value.  Still there are temporal fluctuations of $\langle \op D
(t) \rangle$ as $[\op D,\op H] \ne 0$; see Fig.\
\ref{fig:one_shorttdecay}(b) and also Fig.\ \ref{fig:one}, for
examples.  However, the fluctuations can be quite small as compared
with the time average
\begin{equation}
  \overline{\op D} = \lim_{T\to \infty} \frac{1}{T} \int_{0}^{T}dt \: \langle \op D(t) \rangle \, .
\end{equation}
To quantify these observations, the time average after the initial
decay at short times as well as the relative standard deviation
$(\overline{\op D^{2}} - \overline{\op D}^{2})^{1/2}/\overline{\op
  D}$, as a measure for the temporal fluctuations, are shown as
contour plots in Fig.~\ref{fig:one_stab}.  Some sectional views are
given in Fig.~\ref{fig:sec_one_stab}.

For vanishing couplings $U$ and $V$ the doublon decays on a short-time
scale and is found anywhere in the lattice with a probability of
approximately $0.019$ (for $L=100$) at later times but fluctuations
are strong. For finite and increasing $U$, but keeping $V=0$, the
doublon stability rapidly rises while the relative fluctuations
decrease.  This is understood easily as the energy conservation
described by Eq.\ (\ref{eq:fopt}) becomes strict in the long-time
limit, i.e., a single doublon in an otherwise empty band is completely
stable.\cite{remark}

Using Eq.\ (\ref{eq:doft}), we find the time average for $T\to
\infty$,
\begin{equation}
  \overline{\op D}
  \propto
  \sum_{i}
  \int d\omega A_{i_{0}i_{0},ii}(\omega) A_{ii,i_{0}i_{0}}(\omega) \, ,
  \label{eq:stab}
\end{equation}
to be given by the integrated square of the nonlocal APS spectral
density.  As $A_{ii,i_{0}i_{0}}(\omega)$ consists of a finite number
of $\delta$ peaks for any finite $L$, the integral in Eq.\
(\ref{eq:stab}) is ill defined.  However, one can also compute
$\overline{\op D}$ directly, starting from Eq.\ (\ref{eq:key}),
inserting resolutions of the unity in the form $\ff 1=\sum_{m} |
m\rangle \langle m |$ where $|m\rangle$ is the $m$th eigenstate of
$\op H$.  Assuming the energy spectrum to be non-degenerate and
assuming that there is relaxation at all, we easily find
\begin{equation}
  \overline{\op D}
  =
  \sum_{i} \sum_{m}
  \big| \langle 0 | \ann{d}{i_{0}} | m \rangle \big|^{2}
  \big| \langle 0 | \ann{d}{i} | m \rangle \big|^{2}
  \: .
  \label{eq:stab1}
\end{equation}
With the expressions Eqs.\ (\ref{eq:eom1}) and (\ref{eq:eom2}) for the
corresponding Green's function, and using its Lehmann representation,
this is also seen to be consistent with Eq.\ (\ref{eq:stab}).
Equation (\ref{eq:stab}) provides the long-time ``thermal'' value of
the total double occupancy.

For large $U$, the numerical results of Fig.\ \ref{fig:one_stab} can
be perfectly fitted by
\begin{equation}
  \overline{\op D}
  \simeq 1 - \frac{m}{U^2}  \; ,
  \label{taver}
\end{equation}
with the constant $m>0$ as a parameter. This behavior can be
understood by perturbative arguments using the canonical
transformation discussed in Sec.\ \ref{sec:prop} and Appendix
\ref{sec:effective}: Using the unitary transformation $\op H ' =
e^{i\op S} \op H e^{-i\op S}$ with the generator $\op S$, we find
\begin{equation}
  \langle \op D (t) \rangle  
  =
  \langle 0 | \ann{d}{i_{0}} e^{-i\op S} e^{i\op H't} e^{i\op S} \op D e^{-i\op S} e^{-i\op H't} e^{i\op S} 
  \cre{d}{i_{0}} | 0 \rangle \, .
\end{equation}
Exploiting particle-number conservation, we then get
\begin{equation}
  \langle \op D (t) \rangle  
  =
  \sum_{i} \left| \langle 0 | \ann{d}{i_{0}} e^{-i\op S} e^{i\op H't} e^{i\op S} \cre{d}{i} | 0 \rangle \right|^{2} \, .
\end{equation}
The state
\begin{equation}
  |\psi'_{i}\rangle
  \equiv 
  e^{i\op S} \cre{d}{i} | 0 \rangle 
  =
  \cre{d}{i} | 0 \rangle 
  +
  i [\op S, \cre{d}{i}]
  | 0 \rangle 
  + {\cal O}(J^{2}/U^{2})
\end{equation}
is a linear superposition of a one-doublon and a zero-doublon state
\begin{equation}
  |\psi'_{i}  \rangle
  =
  | \psi'_{i,1} \rangle 
  - \frac{J}{U} 
  | \psi'_{i,0} \rangle 
  + 
  {\cal O}(J^{2}/U^{2}) \,,
\end{equation}
with $| \psi'_{i,1} \rangle = \cre{d}{i} | 0 \rangle$ and $|
\psi'_{i,0} \rangle = - \sum_{k}^{\rm n.n.}  (\cc{k\ua}\cc{i\da} +
\cc{i\ua}\cc{k\da}) | 0 \rangle$.  This characteristic is, separately,
preserved under the time evolution $e^{i\op H't}$.  After some algebra
we find
\begin{eqnarray}
  \langle \op D (t) \rangle  
  &=&
  1+
  \frac{J^{2}}{U^{2}} 
  \sum_{i}
  \langle \psi'_{i,1} | e^{- i\op H_{\rm eff}t} | \psi'_{i_{0},1} \rangle 
  \langle \psi'_{i_{0},0} | e^{i\op H_{\rm eff}t} | \psi'_{i,0} \rangle 
  \nonumber \\
  &+&
  \mbox{H.c.}
  +
  {\cal O}(J^{4}/U^{4})
  \, .
  \label{eq:pertur}
\end{eqnarray}
Hence perturbative in $1/U$ corrections to the total double occupancy
are of the order $J^{2}/U^{2}$.

Equation (\ref{eq:pertur}) furthermore shows that in leading perturbation
order there is a separation of characteristic time scales. The first
matrix element involves energies in the one-doublon subspace and thus
a short-time scale $1/U$.  This is the time scale of the strong
initial oscillations of $\langle \op D (t) \rangle$, seen in Fig.\
\ref{fig:one_shorttdecay}, and once more explains the $1/U$ dependence
of the ``decay time,'' i.e., the position of the first minimum.  The
second matrix element between states in the zero-doublon subspace
provides a longer time scale $\sim 1/J$.  This is the scale on which
the oscillations decay.  Finally, corrections to the $1/U$ scale are
provided by effective doublon-hopping processes.  This results in a
scale $1/J' \sim U/J^{2}$ the effects of which, however, are too weak
to be seen in Fig.\ \ref{fig:one_shorttdecay}.

Finally, let us discuss the results for a finite nearest-neighbor
interaction $V$.  As is shown in Figs.~\ref{fig:one_stab} and
\ref{fig:sec_one_stab}, the initially prepared doublon is most
unstable for $V\approx U$.  Here, the transition $\cre{d}{i}\ket|0>
\ra \cc{i\ua}\cc{i\pm 1,\da}\ket|0>$ becomes resonant.  A further
separation of the fermions beyond nearest-neighbor distances, however,
is the more suppressed the larger $V$ gets. The latter is obvious for
reasons analogous to those given above regarding the $U$ dependencies.
As already noted in the context of propagation patterns above, for
$U=V\neq 0$, an ``extended doublon'' is formed as a linear combination
of a doubly occupied site with states where the two fermions are found
at adjacent sites.  Though the probability for finding two fermions at
the same site anywhere in the lattice shows a minimum for $U=V$ in the
stability map in Fig.~\ref{fig:one_stab}, the one for finding them as
nearest neighbors is almost equally large as can be seen in the
sectional views of the stability map in Fig.~\ref{fig:sec_one_stab}.
Furthermore, the sum of both equals the one for the same value of $U$
but vanishing $V$.  At the same time the oscillations of the double
occupancy as well as the one of nearest-neighbor state exhibit a
maximum (Figs.~\ref{fig:one_stab} and \ref{fig:sec_one_stab}) whereas
their sum does not.  Hence the oscillations cancel each other.

\section{Two doublons}
\label{sec:four-fermions}

\begin{figure}[t!]
  \begin{center}
    \includegraphics[width=0.75\columnwidth]{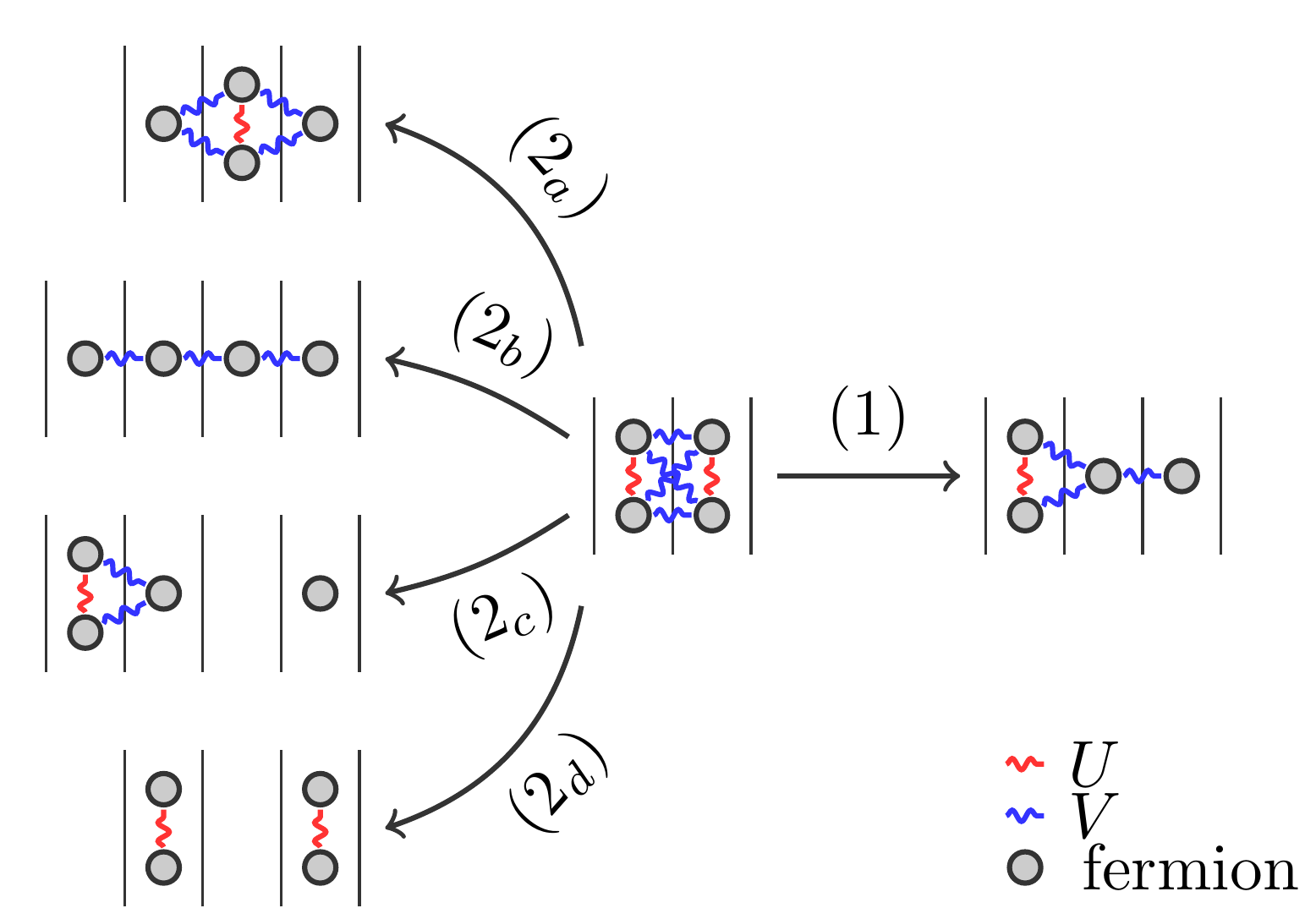}
  \end{center}
  \caption{ Scheme of dominant first-order [$(1)$, right] and
    second-order [$(2)$, left] hopping processes from an initial state
    with two doublons on nearest-neighbor sites to the possible final
    states.  The involved interactions $U$ and $V$ are depicted by
    wiggly lines in red and blue, respectively.  }
  \label{fig:hopproc-nn}
\end{figure}


\begin{figure*}[t!]
  \newcounter{subfignn} \setcounter{subfignn}{0}
  \newcounter{subfignnn} \setcounter{subfignnn}{0}
  \newcounter{subfigsep} \setcounter{subfigsep}{0}

  \renewcommand\thesubfigure{\alph{subfignn}${}_\text{nn}$}%
  \addtocounter{subfignn}{1}%
  \subfloat[$U=10,V=-10$]{\label{fig:two-nnU10V-10}\includegraphics[width=.2\textwidth]{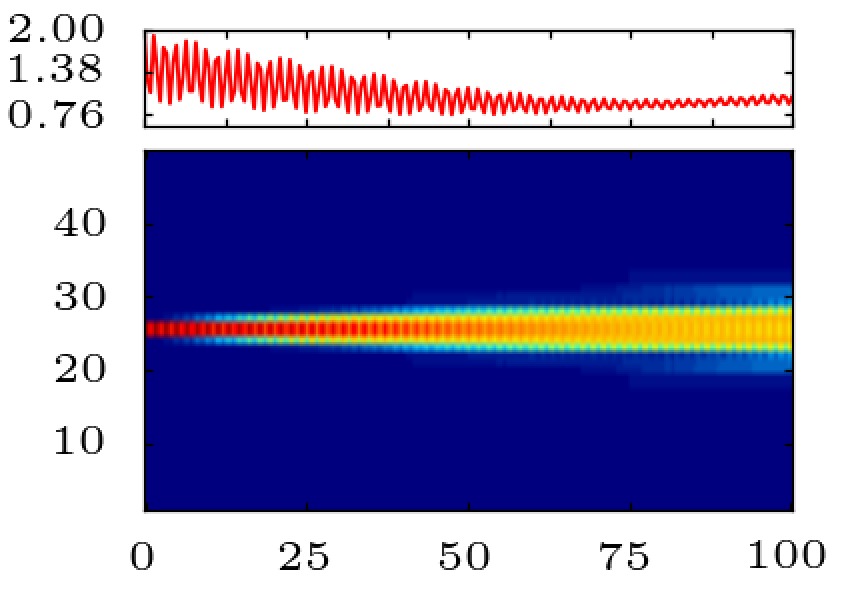}}%
  \addtocounter{subfignn}{1}%
  \subfloat[$U=5,V=-10$]{\label{fig:two-nnU5V-10}\includegraphics[width=.2\textwidth]{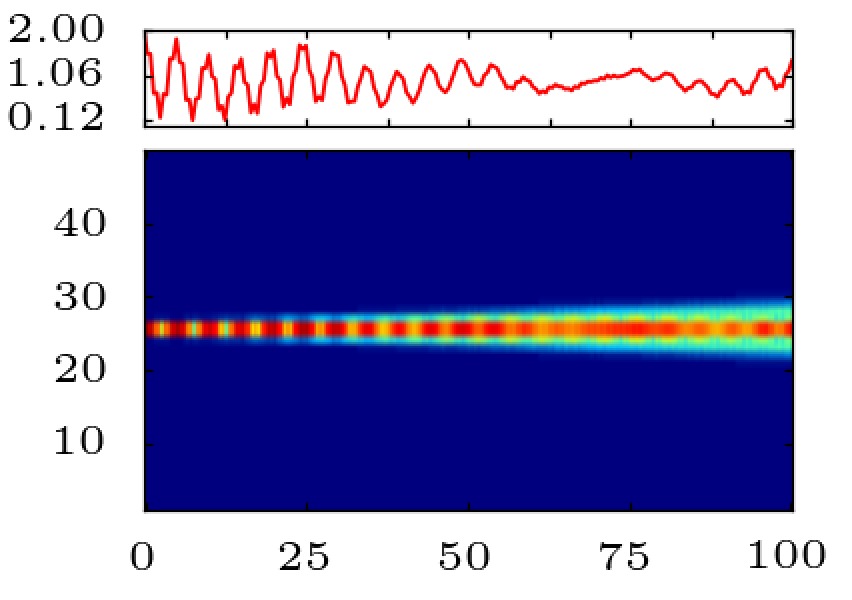}}%
  \renewcommand\thesubfigure{\alph{subfignnn}${}_\text{nnn}$}%
  \addtocounter{subfignnn}{1}%
  \subfloat[$U=5,V=5$]{\label{fig:two-nnnU5V5}\includegraphics[width=.2\textwidth]{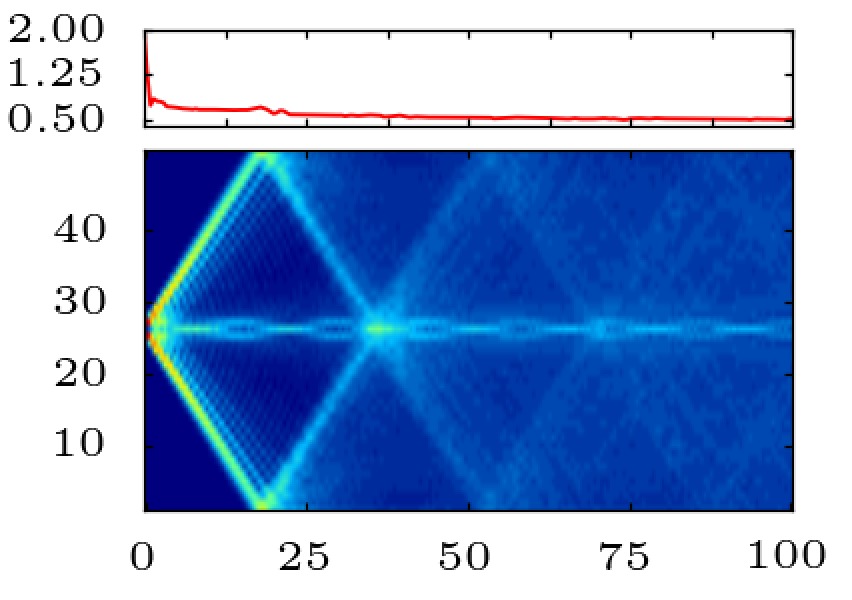}}%
  \addtocounter{subfignnn}{1}%
  \subfloat[$U=10,V=10$]{\label{fig:two-nnnU10V10}\includegraphics[width=.2\textwidth]{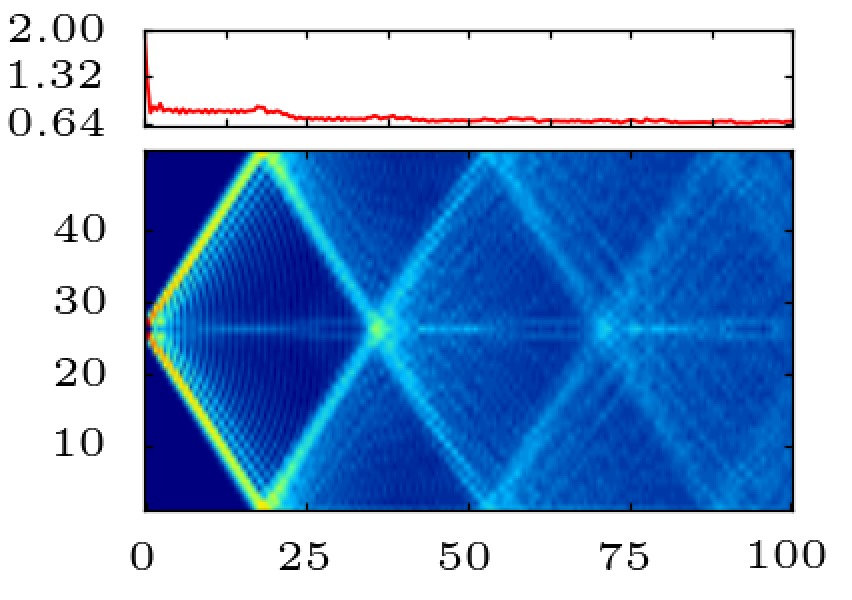}}%
  \renewcommand\thesubfigure{\alph{subfigsep}${}_\text{sep}$}%
  \addtocounter{subfigsep}{1}%
  \subfloat[$U=5,V=5$]{\label{fig:two-sepU5V5}\includegraphics[width=.2\textwidth]{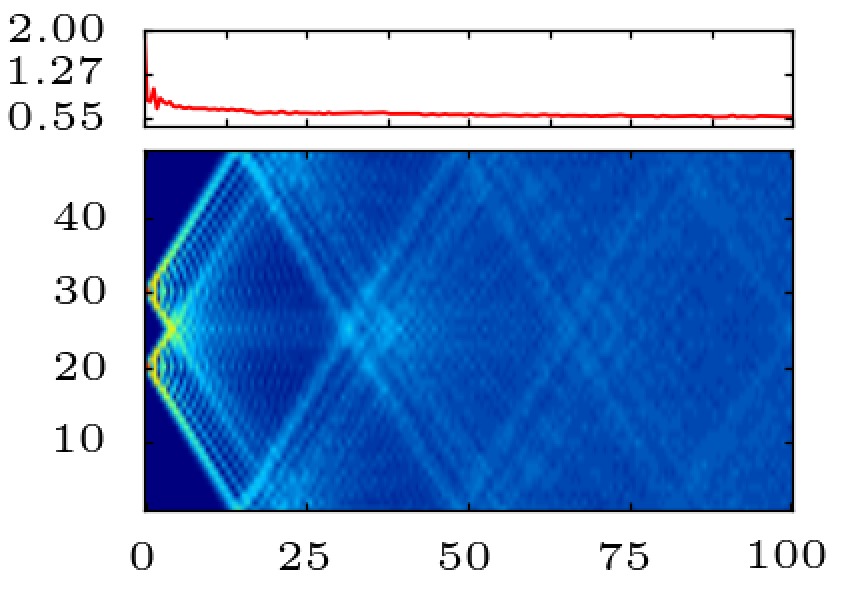}}%
  \\
  \renewcommand\thesubfigure{\alph{subfignn}${}_\text{nn}$}%
  \addtocounter{subfignn}{1}%
  \subfloat[$U=0,V=10$]{\label{fig:two-nnU0V10}\includegraphics[width=.2\textwidth]{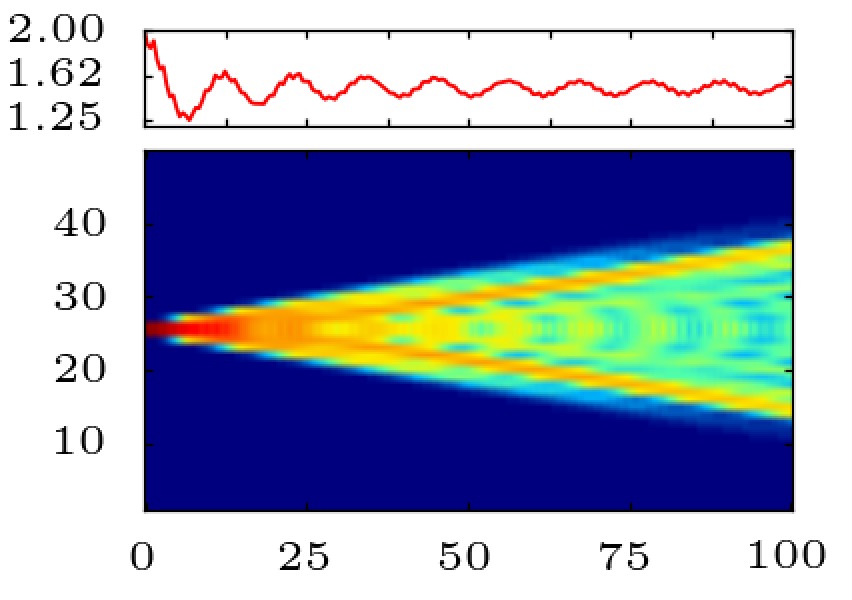}}%
  \addtocounter{subfignn}{1}%
  \subfloat[$U=5,V=0$]{\label{fig:two-nnU5V0}\includegraphics[width=.2\textwidth]{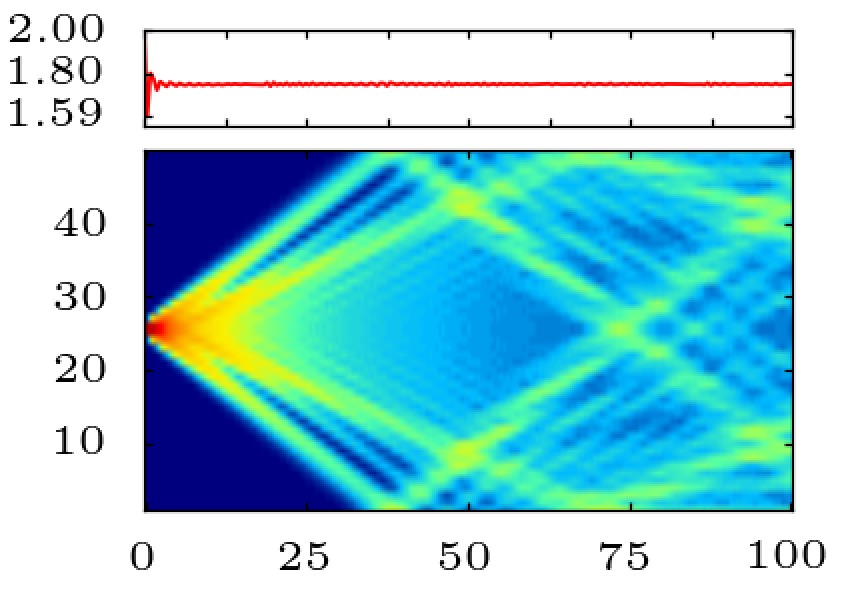}}%
  \renewcommand\thesubfigure{\alph{subfignnn}${}_\text{nnn}$}%
  \addtocounter{subfignnn}{1}%
  \subfloat[$U=0,V=10$]{\label{fig:two-nnnU0V10}\includegraphics[width=.2\textwidth]{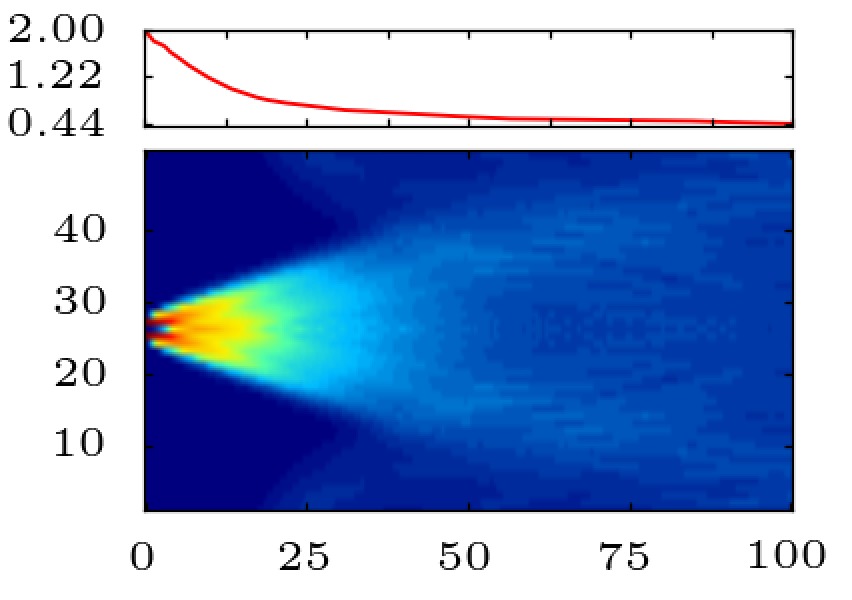}}%
  \addtocounter{subfignnn}{1}%
  \subfloat[$U=6,V=2$]{\label{fig:two-nnnU6V2}\includegraphics[width=.2\textwidth]{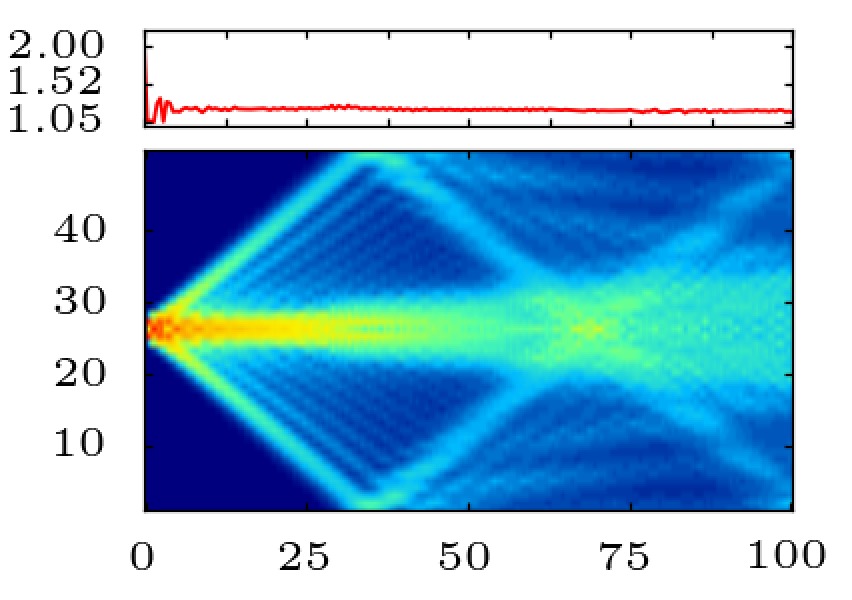}}%
  \renewcommand\thesubfigure{\alph{subfigsep}${}_\text{sep}$}%
  \addtocounter{subfigsep}{1}%
  \subfloat[$U=10,V=5$]{\label{fig:two-sepU10V5}\includegraphics[width=.2\textwidth]{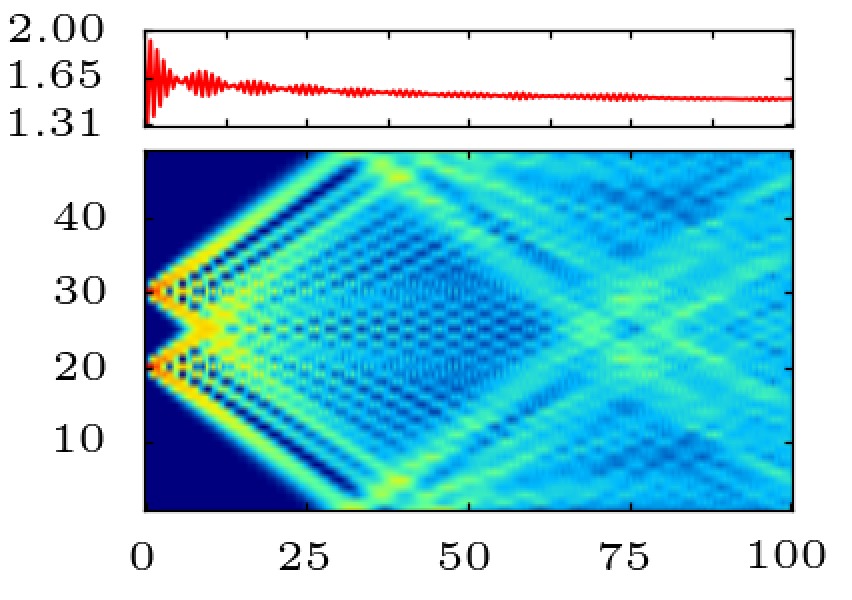}}%
  \\
  \renewcommand\thesubfigure{\alph{subfignn}${}_\text{nn}$}%
  \addtocounter{subfignn}{1}%
  \subfloat[$U=10,V=-5$]{\label{fig:two-nnU10V-5}\includegraphics[width=.2\textwidth]{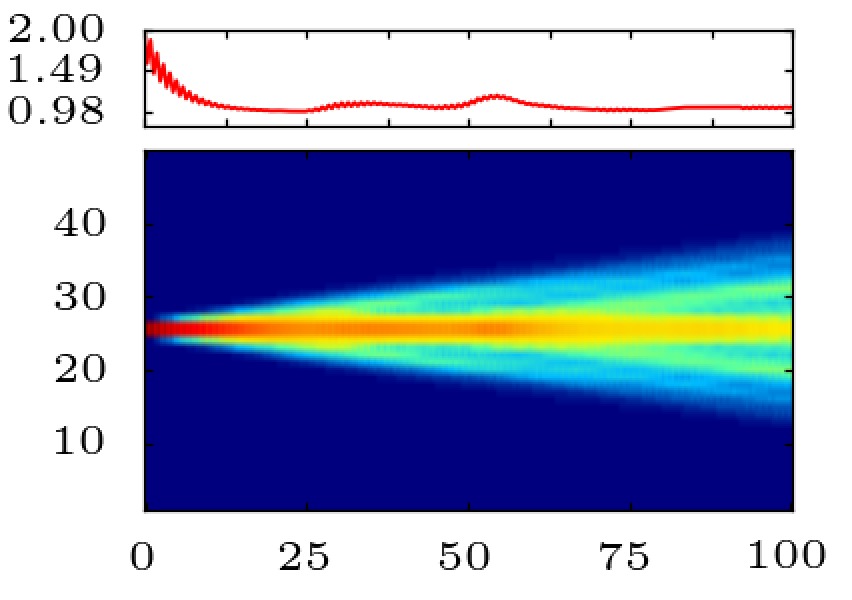}}%
  \addtocounter{subfignn}{1}%
  \subfloat[$U=10,V=10$]{\label{fig:two-nnU10V10}\includegraphics[width=.2\textwidth]{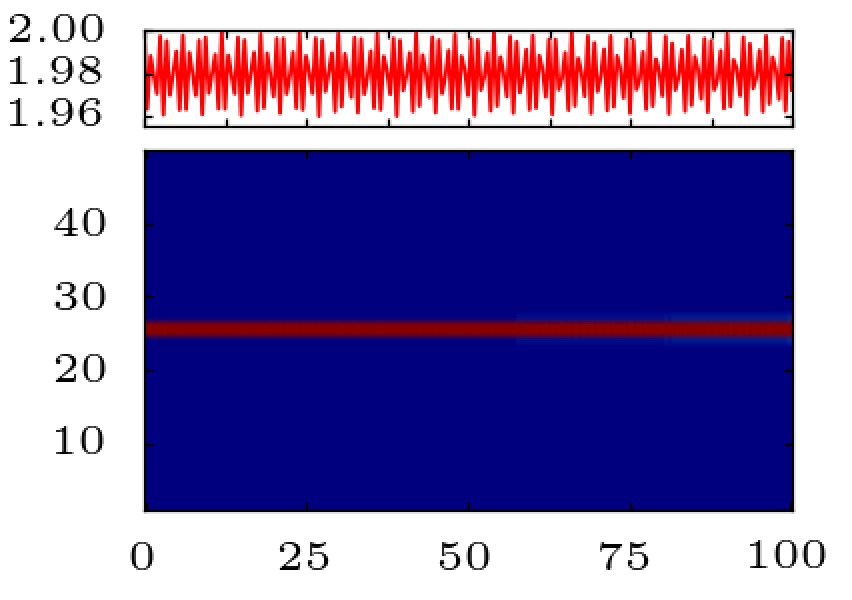}}%
  \renewcommand\thesubfigure{\alph{subfignnn}${}_\text{nnn}$}%
  \addtocounter{subfignnn}{1}%
  \subfloat[$U=8,V=2$]{\label{fig:two-nnnU8V2}\includegraphics[width=.2\textwidth]{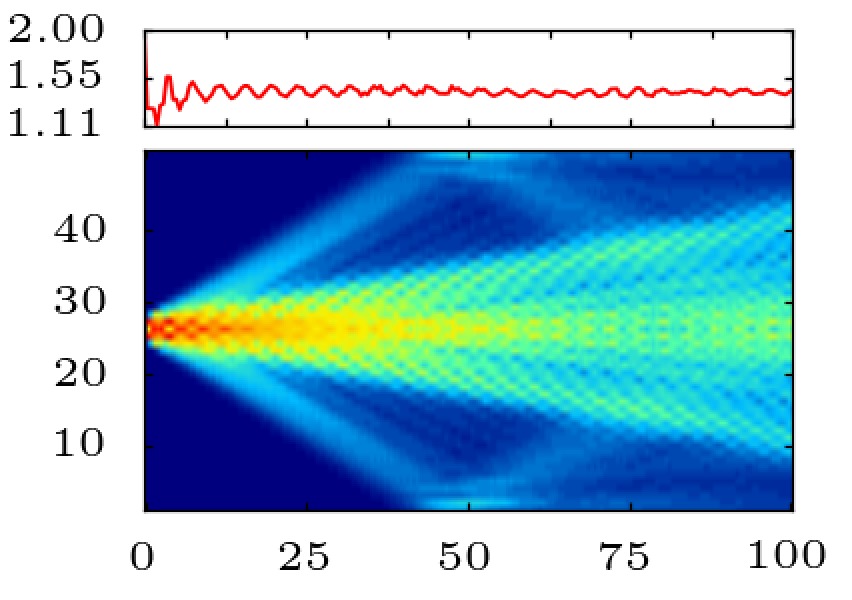}}%
  \addtocounter{subfignnn}{1}%
  \subfloat[$U=8,V=4$]{\label{fig:two-nnnU8V4}\includegraphics[width=.2\textwidth]{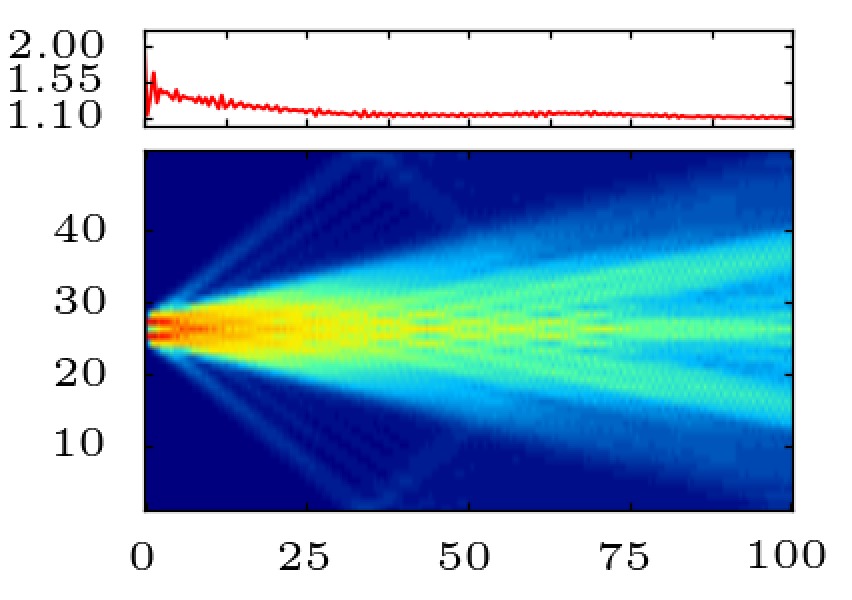}}%
  \renewcommand\thesubfigure{\alph{subfigsep}${}_\text{sep}$}%
  \addtocounter{subfigsep}{1}%
  \subfloat[$U=10,V=0$]{\label{fig:two-sepU10V0}\includegraphics[width=.2\textwidth]{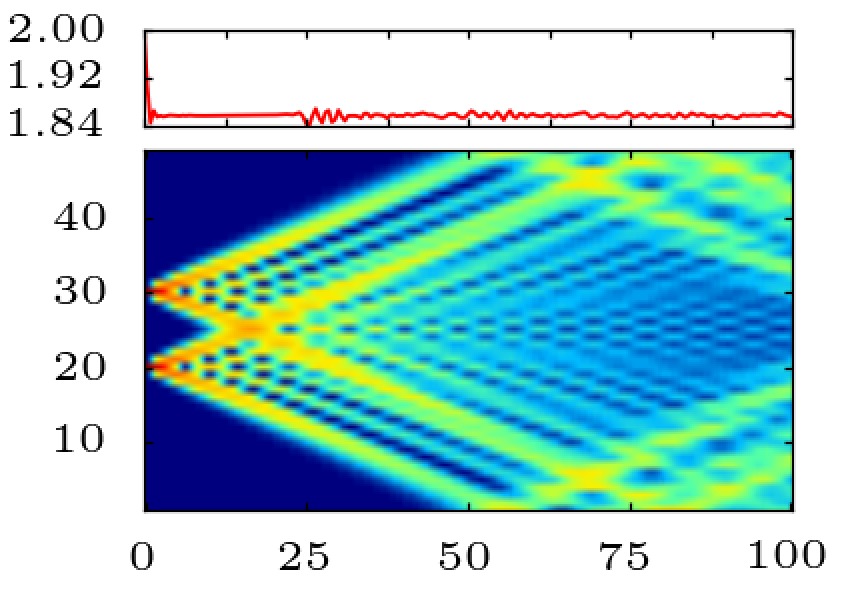}}%
  \caption{ Time dependence of the expectation value of the local
    (main panels) and the total double occupancy (small top panels)
    for different $U$ and $V$ as indicated.  Axis and color code as in
    Fig.~\ref{fig:eff}.  Calculations are performed for different
    initial states as indicated by the bracketed symbols: two pairs
    fermions (two doublons) initially prepared as nearest-neighbors
    (x${}_\text{nn}$), next-nearest-neighbors (x${}_\text{nnn}$) or
    further separated with $|i-j|=10$ (x${}_\text{sep}$) respectively.
    Results for one-dimensional lattice with periodic boundary
    conditions and $L=50,51$ and $49$ sites, respectively.  }
  \label{fig:two}
\end{figure*}

The preceding examinations were restricted to the subspace of two
fermions.  This lacks some important aspects, such as doublon-doublon
and doublon-fermion scattering.  In the following we therefore extend
our study to four-fermion states.  The size of the one-dimensional
lattice is fixed to $L \approx 50$.

\subsection{Initial state with neighboring doublons}

To begin with, consider an initial state at $t=0$ with two doublons at
neighboring sites: $|\psi_\text{ini} \rangle =
\cre{d}{i}\cre{d}{j}\ket|0>$ with $|i-j| = 1$.  In the strong-coupling
limit $U,V \gg J$, this state has a mean energy of the order of $2U+4V
+ {\cal O}(J^{2}/U,J^{2}/V)$: A state with two neighboring doublons
entails two neighboring fermions for each constituent fermion.
Processes starting from this state and involving a single or two
hopping events will dominate the physics in the strong-coupling case
and are sketched in Fig.~\ref{fig:hopproc-nn}.

Figure~\ref{fig:two} shows the time-dependent local and total double
occupancy for different $U$ and $V$.  The overall trends can by
understood by focusing on certain resonant cases as follows.

(i) If $V=-U$ the first order process, referred to as $(1)$ in
Fig.~\ref{fig:hopproc-nn}, becomes resonant: The initial and final
state have the same mean energy up to a small correction of the order
${\cal O}(J^{2}/U,J^{2}/V)$.  In the strong-coupling limit a further
spatial separation of the fermions is suppressed as there is a large
excess energy $U$ or $V$ that cannot be accommodated in the system.  A
propagation of the compound object over many lattice sites is only
possible via second-order hopping processes with a very low
probability as compared to the first-order process (1).  We therefore
expect the two doublons to be basically localized at their initial
positions.  This explains the pattern shown in
Fig.~\subref*{fig:two-nnU10V-10}.

After some settling time the total double occupancy [see top panel in
Fig.~\subref*{fig:two-nnU10V-10}] tends to a value slightly less than
unity which is less than expected for both states that define the
process (1).  We therefore conclude that there is a certain non-zero
probability for the decay of the compound object into fragments
without double occupancy that is not consistent with energy
conservation.  As discussed for the two-fermion case, this is possible
at very short times.

\begin{figure*}[t]
  \subfloat[nearest-neighbors]{\label{fig:two_stab_nn}\includegraphics[width=0.33\textwidth]{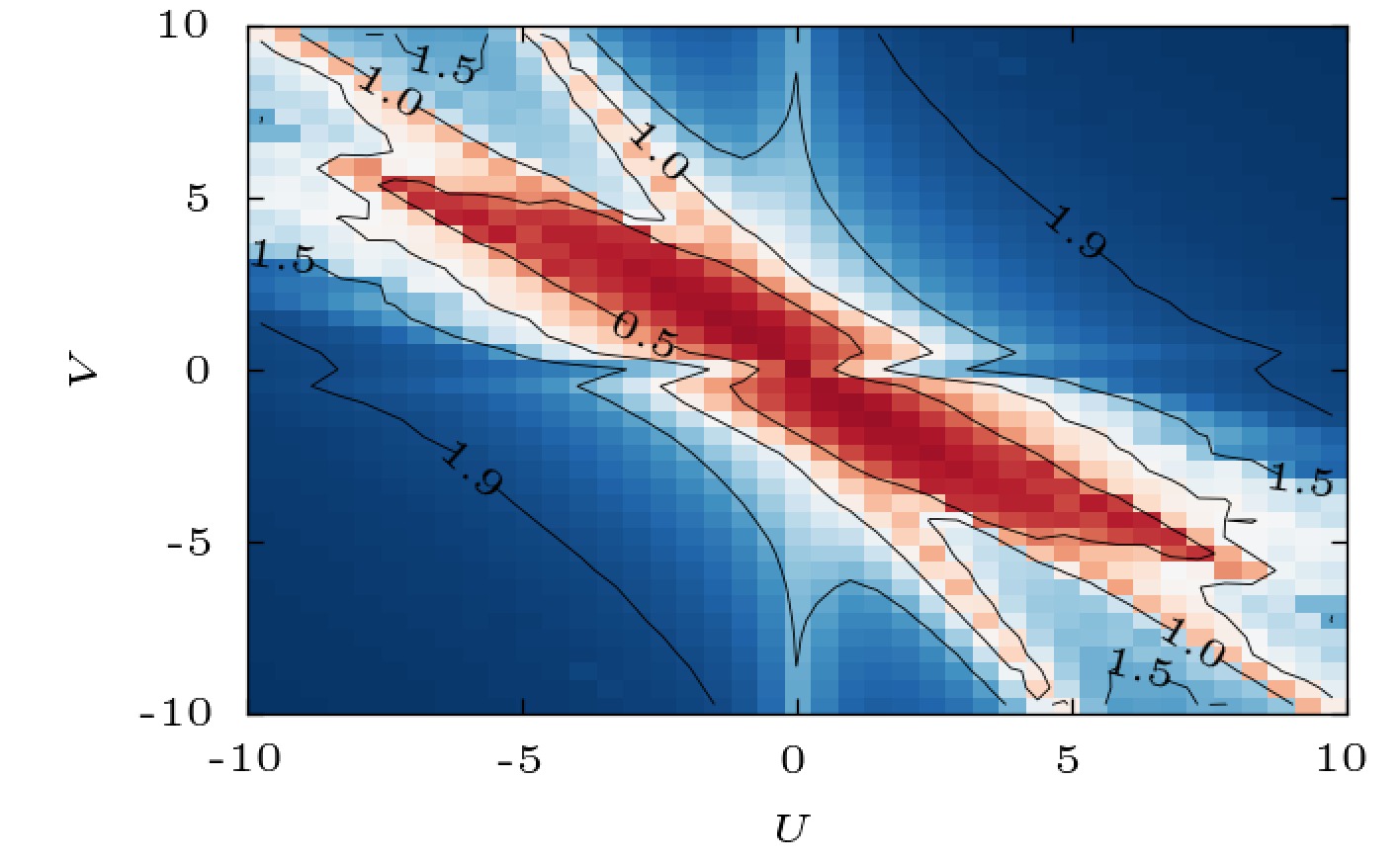}}%
  \subfloat[next-nearest-neighbors]{\label{fig:two_stab_nnn}\includegraphics[width=0.33\textwidth]{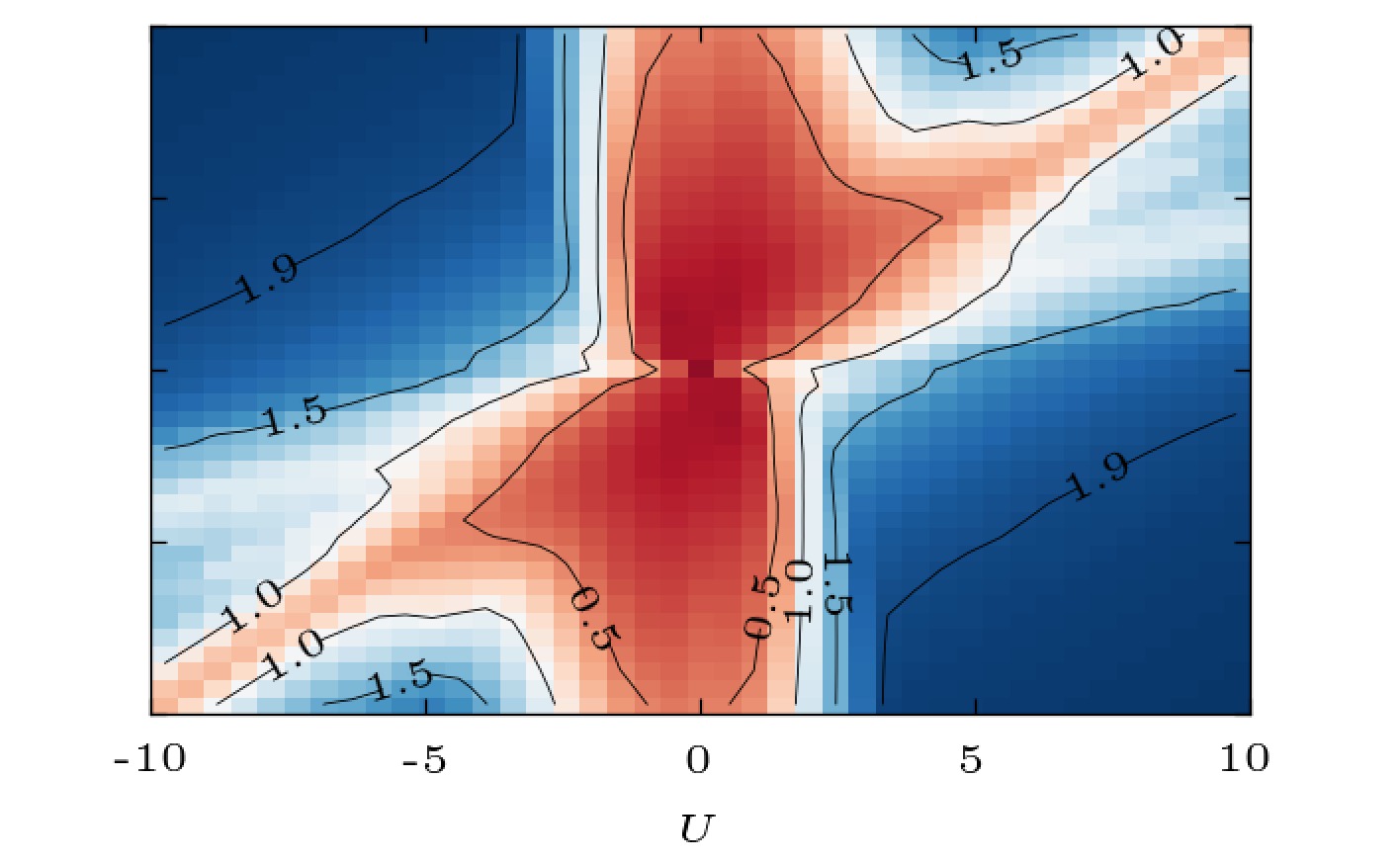}}%
  \subfloat[further
  separated]{\label{fig:two_stab_sep}\includegraphics[width=0.33\textwidth]{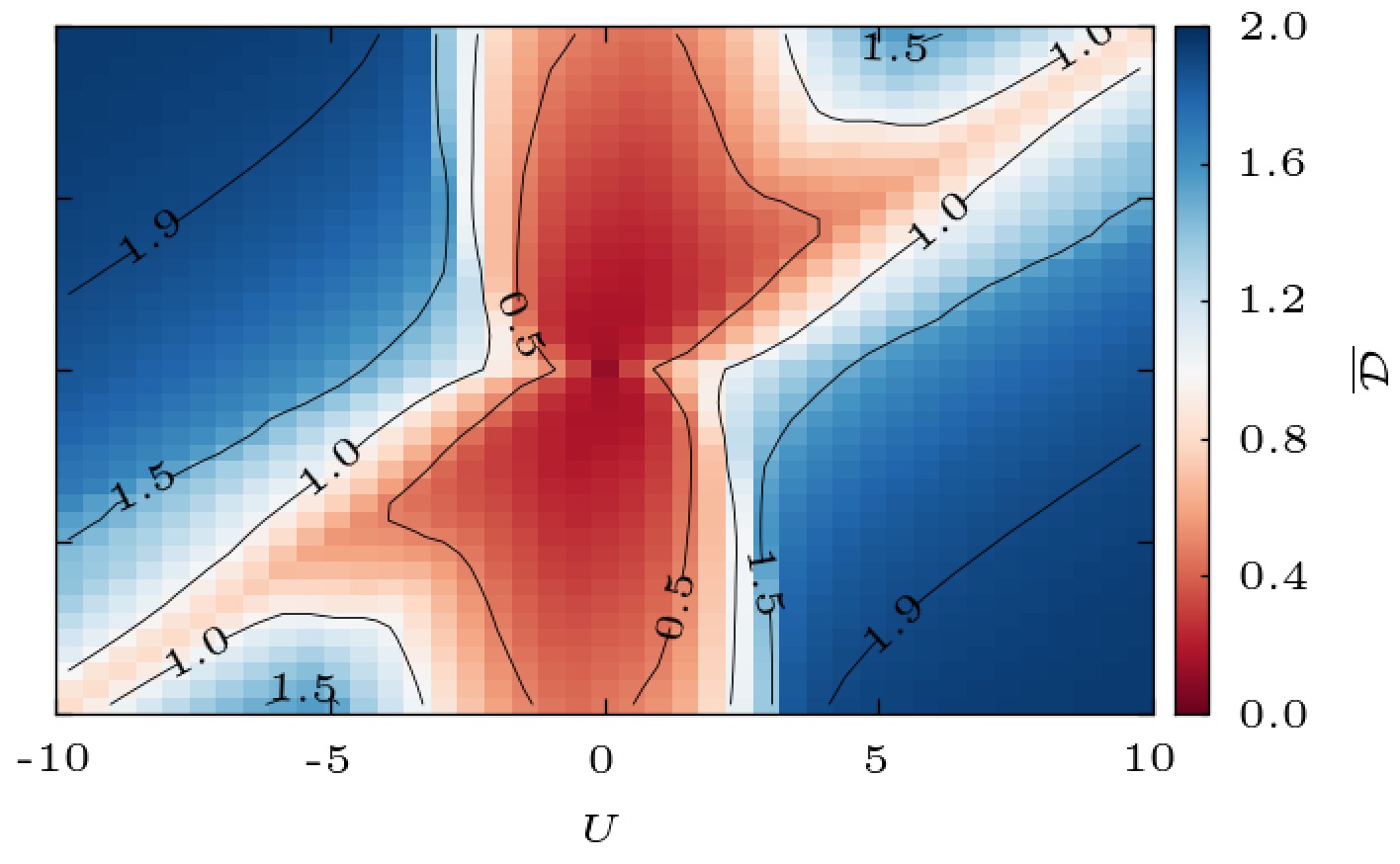}}%
  \caption{ Time average of the total double occupancy $\langle \op
    D(t) \rangle$ for interaction strengths $-10 < U < 10$ and $-10 <
    V < 10$.  Calculations for an initial state with two doublons
    placed at sites $i$ and $j$.  (a) $i$ and $j$ nearest neighbors,
    (b) next-nearest neighbors and (c) $|i-j| = 10$.  Average over the
    time interval $50<t<100$.  The color code is given on the right.
  }
  \label{fig:two_stab}
\end{figure*}

The main dynamical effect, however, consists of a rapid oscillation
between the two states of process (1).  In the map for the time
average $\overline{\op D}$, see Fig.~\ref{fig:two_stab} (left), this
manifests itself as a ``valley'' along the bisecting line of the
second and fourth quadrant.  Furthermore, this is accompanied by a
maximum in the relative fluctuations (not shown), similar as in the
two-fermion case.

(ii) Correspondingly, we find another ``valley'' along the line given
by $V=-2U$ in Fig.~\ref{fig:two_stab} (left).  This is associated with
the second-order process $(2_b)$ in Fig.~\ref{fig:hopproc-nn} which is
resonant here.  Again, there is mainly an oscillation between the two
states of $(2_b)$ which both have the energy $2U+4V = 3V$.  The
process involves a virtual intermediate state with an off-resonant
energy $U+3V$.

As before in case (i), a propagation of the compound object over many
lattice sites is suppressed as it necessarily involves fourth-order
processes.  In fact, Fig.~\subref*{fig:two-nnU5V-10} shows that the
fermions essentially remain close to their initial sites.
 
An oscillation between the two states of $(2_b)$ clearly implies the
total double occupancy to oscillate between approximately $2$ and $0$.
In the long-time limit it tends to relax to a value close to or
slightly less than $1$.

(iii) In case of vanishing $U$, the second-order process $(2_a)$
becomes resonant at the energy $2U+4V = 4V$.  This causes another
branch of minima along the $V$ axis in Fig.~\ref{fig:two_stab} (left).

Opposed to cases (i) and (ii), the four-fermion cluster may propagate
via the $(2_{a})$ process followed by a process inverse to $(2_a)$ but
resulting in two neighboring doublons shifted by one site to the left
or right as compared with the initial state.  Repeated second-order
hopping processes then lead to a more efficient delocalization of the
cluster and thus also of the expectation value for the double
occupancy as is seen in Fig.~\subref*{fig:two-nnU0V10}.

(iv) For a vanishing $V$, the process $(2_d)$ becomes resonant at the
energy $2U+4V = 2U$.  This implies that the initial cluster with two
neighboring doublons can dissociate into two doublons separated at
arbitrarily large distances via second-order hopping processes over
off-resonant intermediate states.  Delocalization is thus very
efficient and results in the pattern displayed in
Fig.~\subref*{fig:two-nnU5V0}.

The propagation pattern is obviously dominated by two ``light cones''
with different velocities.  This can be traced back to the interaction
between the two doublons by comparing with the patterns in
Figs.~\subref*{fig:two-sepU10V5} and \subref*{fig:two-sepU10V0} which
refer to an initial state where the two doublons are well separated
and prepared at a distance $|i-j|=10$ and where the mode with lower
velocity is absent.  It is an open question whether the slow mode is
due to the repulsive hard-core constraint or due to the attractive
interaction in the effective Hamiltonian Eq.\ (\ref{eq:Heff-docc}).
The ``light cone'' associated with the higher velocity is identical to
the one found for propagation of a single doublon, see
Figs.~\subref*{fig:two-nnU5V0} and \subref*{fig:oneU5V0} and mind the
different lattice sizes.

In Fig.~\ref{fig:two_stab} (left), we find a signature of the resonant
process $(2_{d})$ along the $V=0$ line.  As in the two-fermion case,
the doublons are stabilized with increasing $U$.
  
(v) Finally, the process $(2_c)$ gets resonant if $2U + 4V = U + 2V$
which again becomes manifest in a valley, given by
$V=-\frac{1}{2}U$, in the map, Fig.~\ref{fig:two_stab} (left), which
is clearly visible at larger values of $U$ and $V$.

Regarding the mobility, we note that the process $(2_{c})$ can be
either inverted or the fermion triple can move resonantly through the
lattice.  Both possibilities contribute to the propagation pattern
shown in Fig.~\subref*{fig:two-nnU10V-5}.

In all other cases, the initial state shows both a high stability and
a marginal mobility in the strong-coupling limit.
Figure \subref*{fig:two-nnU10V10} gives an example for $U=V=10$.  We
note that the relative fluctuations around the time average amounts to
approximately $1\%$ only.

\subsection{Next-nearest neighbors}

\begin{figure}[b]
  \begin{center}
    \includegraphics[width=.85\columnwidth]{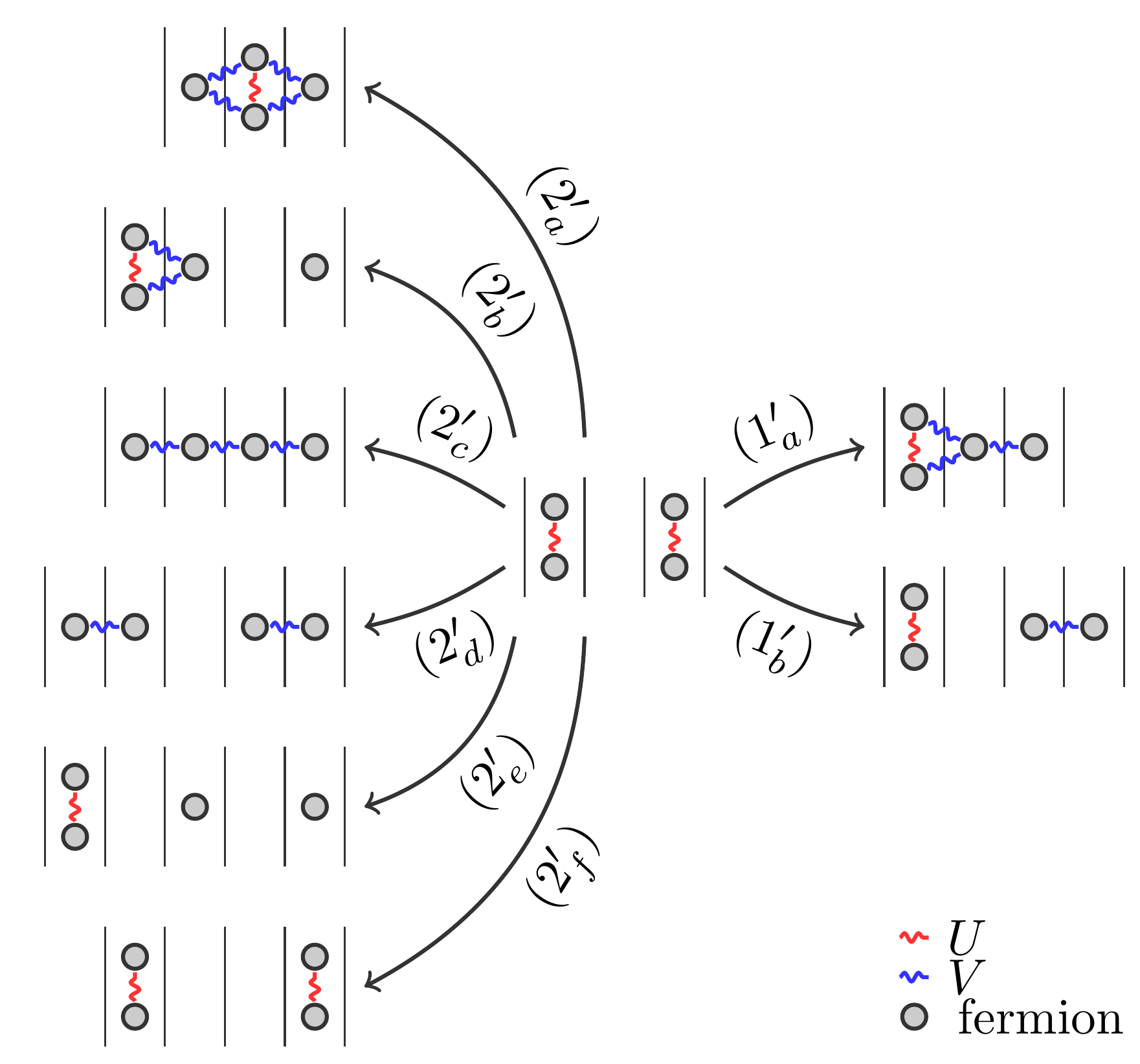}
  \end{center}
  \caption{ Scheme of dominant first-order [$(1')$, right] and
    second-order [$(2')$, left] hopping processes from an initial
    state with two doublons on next-nearest-neighbor sites to the
    possible final states.  The inverse of process $(2_d)$ (see Fig.\
    \ref{fig:hopproc-nn}) is not shown again.  $U$ and $V$ are
    depicted by wiggly lines in red and blue, respectively.  }
  \label{fig:hopproc-nnn}
\end{figure}

Although the underlying physics is the same, the results are
completely different if the two doublons are prepared at sites which
are next-nearest neighbors.  
The calculated propagation patterns are
shown Fig.\ \ref{fig:two} in the third and fourth columns, while
Fig.~\ref{fig:two_stab} (middle) displays the corresponding time
averages.  
The dominant
first-order and second-order hopping processes are sketched in
Fig.~\ref{fig:hopproc-nnn}.  

First, we note that the processes
$(1'_b)$, $(2'_d)$, $(2'_e)$, and $(2'_f)$ are all independent of the
problem's four-particle character.  Provided that the physics is
dominated by those processes, one would expect the propagation pattern
of two initially next-nearest-neighboring doublons to essentially
resemble that of two independent doublons.  In the strong-coupling
limit, this is the case for processes $(1'_b)$, $(2'_d)$ if $U=V$ and
independently of $V$ for $(2'_f)$. As is seen
Figs.~\subref*{fig:two-nnnU5V5} and \subref*{fig:two-nnnU10V10}, the
doublons' propagation is described by about the same maximal effective
hopping as in the case of a single doublon; see Fig.\
\subref*{fig:oneU10V10}, for example.  There is, however, an
additional mode visible in Figs.~\subref*{fig:two-nnnU5V5} and
\subref*{fig:two-nnnU10V10} which results from the two doublons resting
more or less at their initial sites.  This is caused by the respective
inverse hopping processes and basically disappears with increasing
interaction strengths $U=V\to \infty$ and also in the case where the
two doublons are prepared at a larger distance [see Fig.\
\subref*{fig:two-sepU5V5}]. A branch of minima occurs along the line
$U=V$ in the stability map, Fig.~\subref*{fig:two_stab_nnn}, which
looks similar to that obtained in the two-fermion case (cf.\
Fig.~\ref{fig:one_stab}).  The process $(2'_e)$ is resonant only if
$U=0$.  Here the doublons rapidly dissociate into more or less
independent fermions resulting in deep valley around $U=0$ in
Fig.~\subref*{fig:two_stab_nnn}.

The processes $(1'_a)$, $(2'_a)$, $(2'_b)$ and $(2'_c)$ are immanent
to the four-particle character of the problem and become resonant if
$V=\frac{1}{3}U$, $V=\frac{1}{4}U$, $V=\frac{1}{2}U$ or
$V=\frac{2}{3}U$, respectively.  The same holds for the inverse to
process $(2_d)$ (see Fig.\ \ref{fig:hopproc-nn}) which becomes
resonant if $V$ vanishes.  Except for the last one, the doublon number
is changed in all processes.  We therefore expect and find a region of
instability, bounded from below by $V=\frac{1}{4}U$ as can be seen
from the level curves in Fig.~\subref*{fig:two_stab_nnn}. Generally,
the propagation patterns \subref*{fig:two-nnnU6V2},
\subref{fig:two-nnnU8V2} and \subref{fig:two-nnnU8V4} are not easily
interpreted by means of simple perturbative arguments.

It is worth mentioning that for vanishing nearest-neighbor interaction
$V=0$ (not displayed) the doublons essentially show the same spreading
behavior as they did in the case of a single doublon [see
Fig.~\subref*{fig:two-nnU5V0}] and their stability again rises with
$|U|$.  Further, for large couplings of opposite sign $U=-V$, all
processes except for $(2'_f)$ are strongly suppressed.  The patterns
(not displayed) are rather similar to those for a single doublon [see
Fig.~\subref*{fig:two-nnU10V-10}].

\subsection{Further separation in the initial state}

The further away two doublons are prepared in the initial state the
less they influence each other.  We therefore obtain results similar
to those for a single doublon.  
This can be seen from our calculations
with two doublons initially separated by ten sites by comparing, e.g.,
the maps for the long-time averages $\overline{\op D}$,
Figs.~\subref*{fig:two_stab_sep} and \ref{fig:one_stab}, as well
as by comparing the propagation patterns in Figs.~\ref{fig:two} and
\ref{fig:one} for corresponding interaction strengths.

\subsection{Comparison with the bosonic case}

Generally, the propagation patterns considerably differ from the
corresponding ones for doublons formed by bosons.  Motivated by
experiment,\cite{winkler2006repulsivelybound} Petrosyan \emph{et al.}\
\cite{petrosyan2007quantumliquid}
consider the Bose-Hubbard model, $ \op{H} = - \hop \sum_{\left<ij\right>}
\cre{b}{i}\ann{b}{j} + (U/2) \sum_i \n{b}{i} (\n{b}{i} - 1)$, in the
strong-coupling limit with an additional constraint excluding states,
analogous to the Fermi case, with two or more bosons at the same site.
Preparing an initial state with two neighboring doublons, propagation
patterns are obtained which look very similar to our cases $U=-V=10$
or $U=V=10$ [see Figs.~\subref*{fig:two-nnU10V-10} and
\subref*{fig:two-nnU10V10}], i.e., propagation is strongly suppressed.
This can be understood by again referring to a respective effective
model for the strong-coupling limit.  Canonical transformation
yields\cite{petrosyan2007quantumliquid}
\begin{equation}
  \label{eq:bosehubbard-effmodel}
  \op{H}_\text{eff} = \frac{\hop'}{2}
  \sum_{\left<ij\right>} \cre{d}{i}\ann{d}{j} \\ + ( \hop' + U )
  \sum_i \n{d}{i} - 2  \hop' \sum_{\left<ij\right>} \n{d}{i} \n{d}{j} \, .
\end{equation}
Here, $\hat d^{(\dagger)}_i$ denotes the annihilation (creation)
operator for doublons made up of bosons $(\hat b^{(\dagger)})$.  As in
the Fermi case, the effective hopping is given by $ \hop' = 4 \hop^2 /
U$.  Equation\ (\ref{eq:bosehubbard-effmodel}) should be compared with Eq.\
(\ref{eq:Heff-docc}).  In contrast to the fermionic case, the
attractive interaction between two nearest-neighboring doublons is
larger by a factor $4$ for doublons made of bosons.  This explains
the tendency to a strongly suppressed propagation.

It also explains that, in the bosonic case, the formation of clusters
of doublons is favored and phase separation is possible below some
critical
temperature.\cite{petrosyan2007quantumliquid}
Contrary, in the Fermi case, doubly occupied sites may Bose condensate
under certain circumstances.\cite{rosch2008metastable} In fact, we did
not find any indications for a clustering of doublons.  Two doublons
are rather never found to form a bound state unless an explicit
nearest-neighbor interaction $V$ is present.

\section{Doublon-fermion scattering}
\label{sec:df}

The propagation and the decay of a repulsively bound pair is expected
to be strongly affected by the presence of additional fermions.  As a
finite fermion density cannot be studied reasonably by means of the
Krylov approach, we will here consider two additional fermions only.
To this end we first determine the ground state of the Hamiltonian in
the two-fermion subspace $\ket|\Omega_2>$ and subsequently add a
doublon at a certain site $i_{0}$ to define the initial state
$\cre{d}{i_{0}}\ket|\Omega_2>$.  Since the weight of doubly occupied
sites in the ground state is almost vanishing for a lattice with
$L=50$ sites, this setup allows us to study the scattering of the
doublon with almost independently propagating fermions.

\begin{figure}[t]
  \centering
  \includegraphics[width=.7\columnwidth]{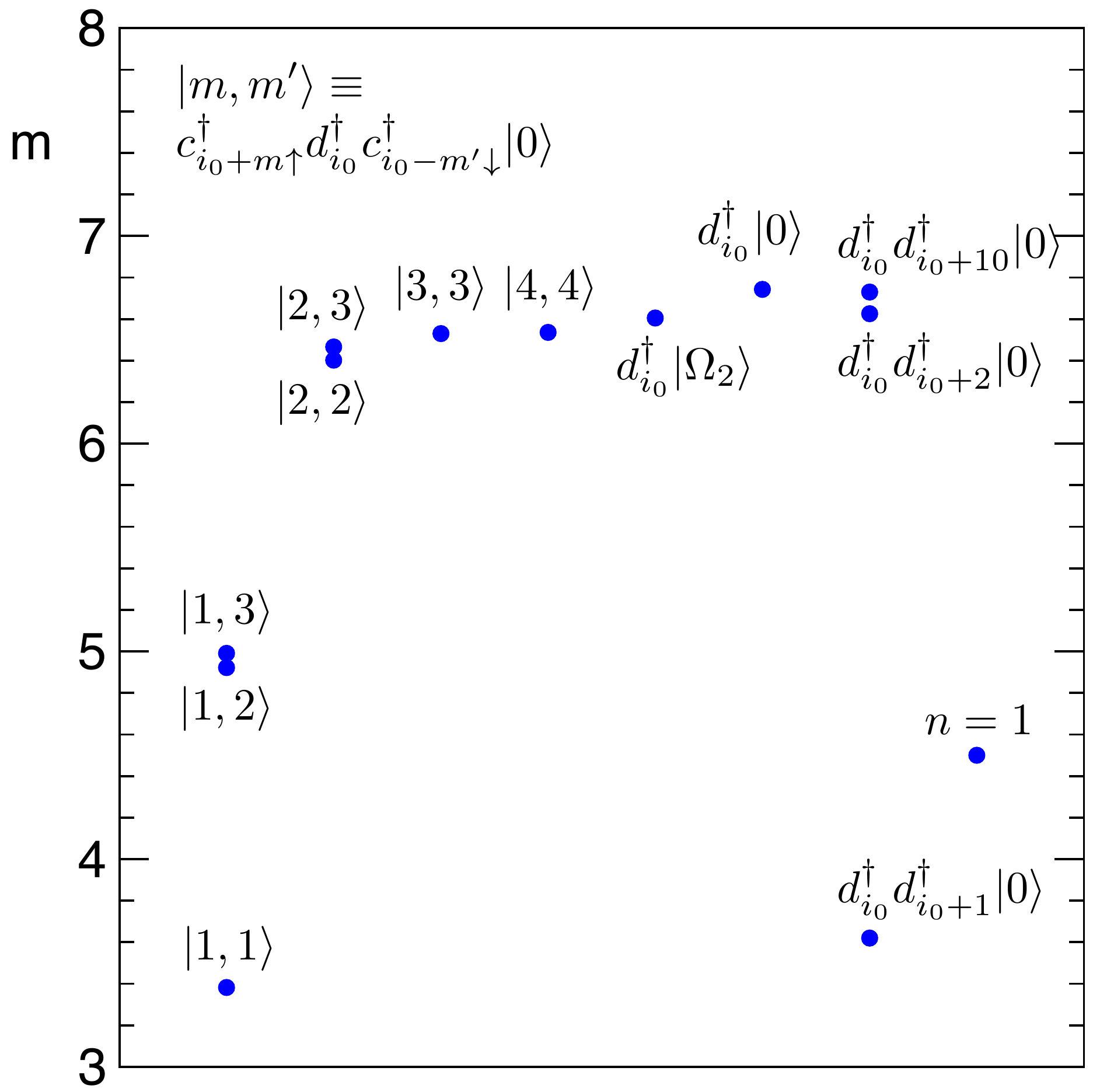}
  \caption{ Coefficient $m$ obtained from a fit of the time-averaged
    expectation value for the total double occupancy $\overline{\op
      D(t)}$ to the observed $U$ dependence: $\overline{\op D(t)}
    \simeq \langle \op D(0) \rangle ( 1 - m / U^{2})$.  $m$ values are
    obtained from linear regression of our data in the range $8\leq U
    \leq 45$.  Results are shown for different initial states as
    indicated and discussed in the text.  $n=1$ refers to the results
    of a time-dependent DMRG calculation by Al-Hassanieh et al., see
    Ref.\ \onlinecite{hassanieh2008excitonsin1dhubbard} and text for
    discussion.  }
  \label{fig:mfit}
\end{figure}

Here we focus on the decay of the doublon for the $V=0$ case only but
consider different initial states.  Besides
$\cre{d}{i_{0}}\ket|\Omega_2>$, we also study the system's time
evolution starting from states where two fermions are prepared at
sites close to the initial position of the doublon $i_{0}$, i.e.\ $|
m, m' \rangle \equiv \cc{i_{0}+m\ua} \cre{d}{i_{0}} \cc{i_{0}-m'\da} |
0 \rangle$.  This is compared to results obtained for two doublons at
nearest-neighboring sites, $\cre{d}{i_{0}} \cre{d}{i_{0}+1} | 0
\rangle$, and two doublons prepared at a distance of 2 and 10, i.e.\
$\cre{d}{i_{0}} \cre{d}{i_{0}+2} | 0 \rangle$ and $\cre{d}{i_{0}}
\cre{d}{i_{0}+10} | 0 \rangle$, respectively.  In all cases we find a
decay of the doublon expectation value on a short-time scale $1/U$
followed by a stabilization to a nearly constant value at large times.
The residual quantum fluctuations are disregarded by looking at the
time average $\overline{\op D}$.  As before, we find that the decayed
doublon fraction scales linearly with $1/U^2$ for large times,
$\overline{\op D(t)} \simeq \langle \op D(0) \rangle ( 1 - m /
U^{2})$.  Hence in the strong-coupling limit the doublon stability is
quantified by the coefficient $m$.  For $m=0$ there is no decay at
all, and a small value for $m$ indicates a rather stable doublon.  Our
results for the different initial states are shown in Fig.\
\ref{fig:mfit}.

Generally, for a system with additional fermions, one expects an
hugely increased phase space for inelastic processes leading to
doublon decay.  On the other hand, the energy-conservation argument
suggests that for strong $U$ a rather complex inelastic process has to
take place to allow for decay, namely a process of high order where a
sufficient number of particles must be involved to dissipate a large
energy of the order of $U$.  While such processes are expected to be
exponentially suppressed for large $U$, they should contribute to some
degree and lead to a destabilization of a doublon.

However, our results for different initial states, as displayed in
Fig.\ \ref{fig:mfit}, just show the opposite trend: The presence of
two additional fermions in the initial state in all cases leads to a
smaller coefficient $m$ in the $1/U^{2}$ decay law.  The strongest
effect is visible for the initial state $| 1,1 \rangle$ where the two
fermions are neighbors of the doublon at $i_{0}$.  Here $m$ is the
smallest and the doublon is most stable.  $m$ increases with
increasing distance of one of the fermions from the position of the
doublon; see the initial states $|1,2\rangle$ and $|1,3\rangle$.  It
further increases if also the second fermion is positioned at a
distance from $i_{0}$ (see $|2,2\rangle$, $|3,3\rangle$, and
$|4,4\rangle$), and it approaches the value obtained for the case
where both fermions are delocalized in the ground state
$\cre{d}{i_{0}}\ket|\Omega_2>$.  The maximum value is obtained for the
isolated doublon in an otherwise empty lattice, i.e., for
$\cre{d}{i_{0}} | 0 \rangle$.  If the two fermions themselves form a
second doublon, see the results for $\cre{d}{i_{0}} \cre{d}{i_{0}+x} |
0 \rangle$ in Fig.\ \ref{fig:mfit}, this again tends to stabilize the
original one: $m$ decreases with decreasing distance $x$ between the
two doublons.

These trends can be understood if the doublon dynamics is considered
at short times: First-order-in-$J$ time-dependent perturbation theory
shows that doublon decay is allowed on a time scale $1/U$ as has been
detailed in Sec.\ \ref{sec:decay}.  Here, one can argue that an
unoccupied site neighboring the doublon is necessary for the decay
process as the immediate surrounding is relevant for its start.
Hence, the coefficient $m$ is the smaller and the doublon is more
stable if decay channels are blocked by localized fermions or doublons
close to the doublon at $i_{0}$ and, to a lesser extent and depending
on the size of the lattice, even by two delocalized fermions in the
two-fermion ground state.  This nicely explains the results described
above.

After that time scale, energy conservation as expressed by Fermi's
golden rule, applies and the total double occupancy virtually relaxes
to a constant value.  As analyzed in Sec.\ \ref{sec:decaylong}, the
probability for the dissociation of a doublon should then scale as
$1/U^2$.  On an for large $U$ extremely long time scale, which
exponentially depends on
$U$,\cite{strohmaier2010observationdoublondecay} contributions from
higher-order perturbation theory in $J/U$ become important and would
generally allow for further decay in more complex
processes.\cite{hansen2011splithubbardbands}

In this context it is interesting to compare our results with the
those of a time-dependent density-matrix renormalization-group (DMRG)
study by Al-Hassanieh \emph{et al.} \cite{hassanieh2008excitonsin1dhubbard}
where the decay of a doublon created by a nearest-neighbor
particle-hole excitation of a half filled one-dimensional Fermi
Hubbard model was considered.  The DMRG calculations show (i) a fast
decay at a characteristic time scale $1/U$, (ii) a basically constant
double occupancy at larger times up about 40$J^{-1}$, and (iii) a
$1/U^{2}$ scaling of the decayed fraction of the doublon.  All this
agrees perfectly with our results obtained for four fermions only.
The $m$ coefficient taken from the DMRG results
\cite{hassanieh2008excitonsin1dhubbard} is also included in Fig.\
\ref{fig:mfit} (``$n=1$'') and is found to be close to that obtained
for the $|1,2\rangle$ initial state.  Even this is plausible since the
spin-dependent site occupations of the state $|1,2\rangle$ and of the
initial state of the DMRG calculation are the same in the immediate
environment of $i_{0}$.

The at least qualitative agreement with the dynamics of the
half filled model on the time scale accessible to time-dependent DMRG
appears as surprising at first sight: Clearly, the initial local
blocking of decay channels is the same in the four-fermion and in the
half filled case but this would only explain an agreement on a time
scale much shorter than the one accessible by DMRG.  We suggest that
it is important to take into account an additional argument here,
namely, the fact that decay of the doublon on intermediate time scales
larger than $1/U$ is basically ruled out by energy conservation while
on time scales shorter than $1/U$ it is only the immediate surrounding
of the doublon that counts.  This would explain the almost
quantitative agreement with the DMRG results of Ref.\
\onlinecite{hassanieh2008excitonsin1dhubbard}.

On the other hand, this argument leaves the possibility for an, e.g.,
exponential-in-$t$ decay law on much larger time
scales. \cite{hansen2011splithubbardbands,strohmaier2010observationdoublondecay}
This might be expected on general grounds as adding more degrees of
freedom to the system should strongly increase the phase space
available for decay in energy-conserving processes where the doublon
energy $U$ is dissipated to a large number of particle-hole or spin
excitations.  Those processes, however, require a huge time scale to
contribute significantly to the doublon decay, possibly well beyond
the time scales accessible by DMRG.

Note that a quantitative comparison with the DMRG study of Ref.\
\onlinecite{enss2011lightconerenorm} for the half filled Hubbard model
is not possible, as an initial state where doubly occupied and empty
sites alternate is considered there.  Still the qualitative features
are rather similar.


\section{Summary}
\label{sec:summary}

Concluding, the real-time dynamics of two or a few more strongly
interacting Fermions moving in a periodic lattice potential exhibits a
surprisingly rich physics which is not only linked to experiments with
ultracold atoms trapped in optical lattices but also to electron
spectroscopy of metal surfaces as well as to rather general questions
on the propagation and decay of bound quantum states and the
relaxation of quantum systems prepared in a highly excited initial
state.  
Here we have employed a Krylov-space method 
\cite{lanczos1950iteration,park1986unitary,saad1992analysiskrylovsubspaceapprox,hochbruck1997krylov,hochbruck1999exponentialintegrators,molervanloan2003expmatrix,manmana2005timev1d}
to study few-particle 
systems with a moderately large Hilbert-space dimension.
Even the analysis of the two-fermion case helps us to
understand important concepts such as the temporal stability of a
doublon, i.e., a repulsively bound pair of fermions.  

The decay of a doublon in an otherwise empty system is possible on a very-short time
scale $1/U$ where energy conservation, within the spirit of
time-dependent first-order perturbation theory,
does not apply.  Using perturbative diagonalization of the Hamiltonian
by means of a canonical transformation, one can understand the
observed $1/U^{2}$ dependence of the fraction of the doublon that has
decayed in the long-time limit.  

The time average of the total double
occupancy is found to be given by a quantity defined for the
equilibrium or ground state of the system, namely the integrated
square of the spectral density related to appearance-potential
spectroscopy.  But also the fully time-dependent local double
occupancy can be expressed in terms of this spectral function, which
must be seen as an unexpected interrelation valid for a two-particle
system only.

The spatiotemporal evolution of the expectation value of the local
double occupancy can be understood by perturbative arguments, 
even in the case of a non zero nearest-neighbor interaction $V$.
In the case of four fermions, the propagation patterns are much more
complicated.  
Still, we could demonstrate that the real-time dynamics after preparation 
of different initial states can be understood in most but not all cases 
by perturbative arguments.

The physics of a finite density of doublons consisting of fermions is
known to be rather different from the case of doublons made of bosonic
particles which undergo a transition to a phase-separated state
instead of Bose
condensation.\cite{petrosyan2007quantumliquid,rosch2008metastable}
Consistent with this, we did not find any indications for a clustering
of doublons consisting of fermions unless an explicit
nearest-neighbor-interaction $V$ is present.

Surprisingly, there is a rather regular trend concerning the decay of
a single doublon in the presence of two more fermions.  
The total double occupancy, apart from quantum fluctuations, relaxes
to a constant value after an initial decay on a time scale $1/U$, and the
long-time average deviates from the initial value by a fraction that
scales with $U$ as $1/U^{2}$ in the strong-coupling limit, like in the
case where there are no additional fermions, but with a coefficient $m$
that characteristically depends on the initial state.  

$m$ is found to decrease and thus the stability of the doublon is found to 
increase when two fermions are added --- a result which at first sight is
conflicting with the expectation that adding more degrees of freedom
to the system should strongly increase the phase space available for
decay in energy-conserving processes where the doublon energy $U$ is
dissipated to a large number of particle-hole or spin excitations.
Those processes, however, require a huge time scale to contribute
significantly to the doublon decay.  More important for the stable
fraction of the doublon is the local environment in the initial state
as the main effect of an additional doublon or of additional fermions
in its vicinity is to block decay channels on the short-time scale on
which decay is possible rather than ruled out by energy conservation.
This is a general argument which apparently also applies to the
half filled case, for example.  In fact, we find almost quantitative
agreement with a time-dependent DMRG
calculation.\cite{hassanieh2008excitonsin1dhubbard} 
On the other hand,
the argument leaves the possibility for an, e.g., exponential-in-$t$
decay law on much larger time scales which might be expected on
general
grounds.\cite{hansen2011splithubbardbands,strohmaier2010observationdoublondecay}


\acknowledgments

We would like to thank H.\ Moritz and M.\ Eckstein for helpful
discussions.  The work was supported by the Deutsche
Forschungsgemeinschaft within the Sonderforschungsbereich 925 (project
B5).

\appendix

\section{Krylov approach}
\label{sec:krylov}

For a given vector $u$ the $n$th Krylov subspace of the full Hilbert
space is defined by \cite{krylov1931numericalsolution}
\begin{align}
  \vs K_n(u,\op H) := \operatorname{span} \left\{ u, \op H u, \dots,
    \op H^{n-1} u \right\} \,.
  \label{eq:def-krylovspace}
\end{align}
Typically, the Krylov-space dimension $n \ll d$.  An orthogonal basis
of $\vs K_n$ can be obtained efficiently via the Lanczos recursion
formula\cite{lanczos1950iteration}
\begin{equation}
  u_{k+1} = \op H u_k - a_k u_k - b_k^2 u_{k-1}
  \qquad ( k = 0,\dots, n-1 ) \,, \label{eq:lanczos-rec}
\end{equation}
with the coefficients $a_k = \langle u_k | \op H u_k \rangle / \langle
u_k | u_k \rangle $ and $b_k^2 = \langle u_k | u_k \rangle / \langle
u_{k-1} | u_{k-1} \rangle$ and the initial values $b_0 = 0$ and
$u_{-1} = 0$.  In the normalized Lanczos basis $\{v_i\}$, with $v_i =
u_i / \| u_i \|$, the Hamiltonian is represented by a tridiagonal
matrix $T$ with diagonal elements $a_0,\dots,a_{n-1}$ and secondary
diagonal elements $b_1,\dots,b_{n-1}$.  Hence we can write $T =
V^\dagger \op H V$, where the matrix $V$ is made up by the basis
vectors $v_i$, i.e., $V = \left( v_0, \dots, v_{n-1} \right)$.

The time evolution of a state $\psi(t) \in \vs K_n = \vs K_n(t)$
approximates its time evolution in the whole Hilbert space:
$\psi(t+\Delta t) \approx V e^{-i T (t+\Delta t)} V^\dagger \psi(t)$.
Here $\psi(t)$ is chosen to be the start vector of the Lanczos
recursion [Eq.~\ref{eq:lanczos-rec}], i.e., the Krylov space at time
$t$ is adjusted to the system's state at $t$.  For a given small time
step $\Delta t$, the approximation can be controlled to a high
accuracy by adjusting the Krylov-space dimension.  Longer time
evolutions are carried out successively by using the propagated state
as the new initial state and adapting $T$ and $V$ after each Lanczos
time step.  It is important to note that this kind of approximation
preserves the unitarity of the time evolution.

Since the diagonalization of the fairly small $n\times n$ matrix $T$
is numerically cheap, the computational effort is dominated by the
$n-1$ matrix-vector multiplications that are necessary to construct
the Lanczos basis and by the number of time steps.  In this work we
dealt with Hilbert spaces with $d = 10^4 \dots 10^6$ dimensions.  For
calculations where, e.g., 200 time steps $\Delta t = 0.5$ are
performed, highly accurate results are obtained using Krylov spaces
with less than $n\approx 20$ dimensions only.

\section{Effective low-energy model}
\label{sec:effective}
We consider the Hamiltonian, Eq.\
(\ref{eq:hubbard-model}), for $V=0$ in the strong-coupling limit $U\gg
J$.  The goal is to perturbatively derive an effective low-energy
Hamiltonian preserving the total double occupancy.  This is done
employing the method of canonical transformations (see also Refs.\
\onlinecite{petrosyan2007quantumliquid} and 
\onlinecite{rosch2008metastable}).

First, the hopping term $\Hhop$ is subdivided into parts preserving or
changing the total double occupancy of the system. Expressing the
identity by number operators for particles and holes, namely $\ff
1_{i\sigma} = \hc{i\sigma} + \nc{i\sigma}$, one may write
\begin{align}
  \Hhop &= -\hop \sum_{\left<ij\right>}\sum_{\sigma} \left(
    \nc{i\bar\sigma}\cc{i\sigma}\ca{j\sigma}\nc{j\bar\sigma} +
    \hc{i\bar\sigma}\cc{i\sigma}\ca{j\sigma}\hc{j\bar\sigma}
  \right) \notag \\
  &\hphantom{=} -\hop \sum_{\left<ij\right>}\sum_{\sigma}
  \nc{i\bar\sigma}\cc{i\sigma}\ca{j\sigma}\hc{j\bar\sigma} -\hop
  \sum_{\left<ij\right>}\sum_{\sigma}
  \hc{i\bar\sigma}\cc{i\sigma}\ca{j\sigma}\nc{j\bar\sigma} \notag \\
  &=: \Hhop^0 + \Hhop^+ + \Hhop^-\,, \label{eq:Hhop0+-}
\end{align}
where the double occupancy is raised/lowered by $\Hhop^\pm$ and
preserved by $\Hhop^0$, since
\begin{equation}
  \label{eq:hamilton-commutators}
  \com[\op H_U,\Hhop^\nu] = \nu U \Hhop^\nu\,, \qquad \nu \in \{0,\pm \}\,.
\end{equation}
The unitary transformation is performed perturbatively:
\begin{equation}
  \label{eq:canonical-transf}
  \op H ' = e^{i\op S} \op H e^{-i\op S} \approx \op H + i \com[\op
  S, \op H] + \frac{i^2}{2} \com[\op S,{\com[\op S,\op H]}] + \dots \: .
\end{equation}
$\Hhop^\pm$ can be eliminated by choosing the generator to be $\op S =
-\frac{i}{U} \left( \Hhop^+ - \Hhop^-\right)$.  Up to order
$\hop^2/U$, we end up with the effective model
\begin{equation}
  \op H_{\text{eff}} = \Hhop^0 + \op H_U + \frac{1}{U}
  \com[\Hhop^+,\Hhop^-]  \,, \label{eq:Heff} 
\end{equation}
which, besides the total particle number, conserves the total double
occupancy in addition.  We can therefore restrict ourselves to a
system without any singly occupied site.  Exploiting this fact, Eq.\
(\ref{eq:Heff}) takes, after some straightforward algebra, the form
given by Eq.\ (\ref{eq:Heff-docc}).


%

\end{document}